\documentclass[usenatbib]{mn2e}
\usepackage{graphicx}
\usepackage{times}

\newcommand\DMS{DMS12}

\begin{document}

\title[Spectroscopic Confusion]{Spectroscopic Confusion: Its Impact on Current and Future Extragalactic HI Surveys}
\author[Jones et al.]{Michael G. Jones$^{1}$\thanks{E-mail: jonesmg@astro.cornell.edu}, Emmanouil Papastergis$^{2}$, Martha P. Haynes$^{1}$ \newauthor and Riccardo Giovanelli$^{1}$ 
\\
$^{1}$Center for Radiophysics and Space Research, Space Sciences Building, Cornell University, Ithaca, NY 14853, USA
\\
$^{2}$Kapteyn Astronomical Institute, University of Groningen, Landleven 12, Groningen NL-9747AD, Netherlands}

\maketitle

\begin{abstract}
We present a comprehensive model to predict the rate of spectroscopic confusion in HI surveys, and demonstrate good agreement with the observable confusion in existing surveys. Generically the action of confusion on the HI mass function was found to be a suppression of the number count of sources below the `knee', and an enhancement above it. This results in a bias, whereby the `knee' mass is increased and the faint end slope is steepened. For ALFALFA and HIPASS we find that the maximum impact this bias can have on the Schechter fit parameters is similar in magnitude to the published random errors. On the other hand, the impact of confusion on the HI mass functions of upcoming medium depth interferometric surveys, will be below the level of the random errors. In addition, we find that previous estimates of the number of detections for upcoming surveys with SKA-precursor telescopes may have been too optimistic, as the framework implemented here results in number counts between 60\% and 75\% of those previously predicted, while accurately reproducing the counts of existing surveys. Finally, we argue that any future single dish, wide area surveys of HI galaxies would be best suited to focus on deep observations of the local Universe ($z < 0.05$), as confusion may prevent them from being competitive with interferometric surveys at higher redshift, while their lower angular resolution allows their completeness to be more easily calibrated for nearby extended sources.
\end{abstract}


\section{Introduction}

\begin{figure*}
\centering
\includegraphics[width=\columnwidth]{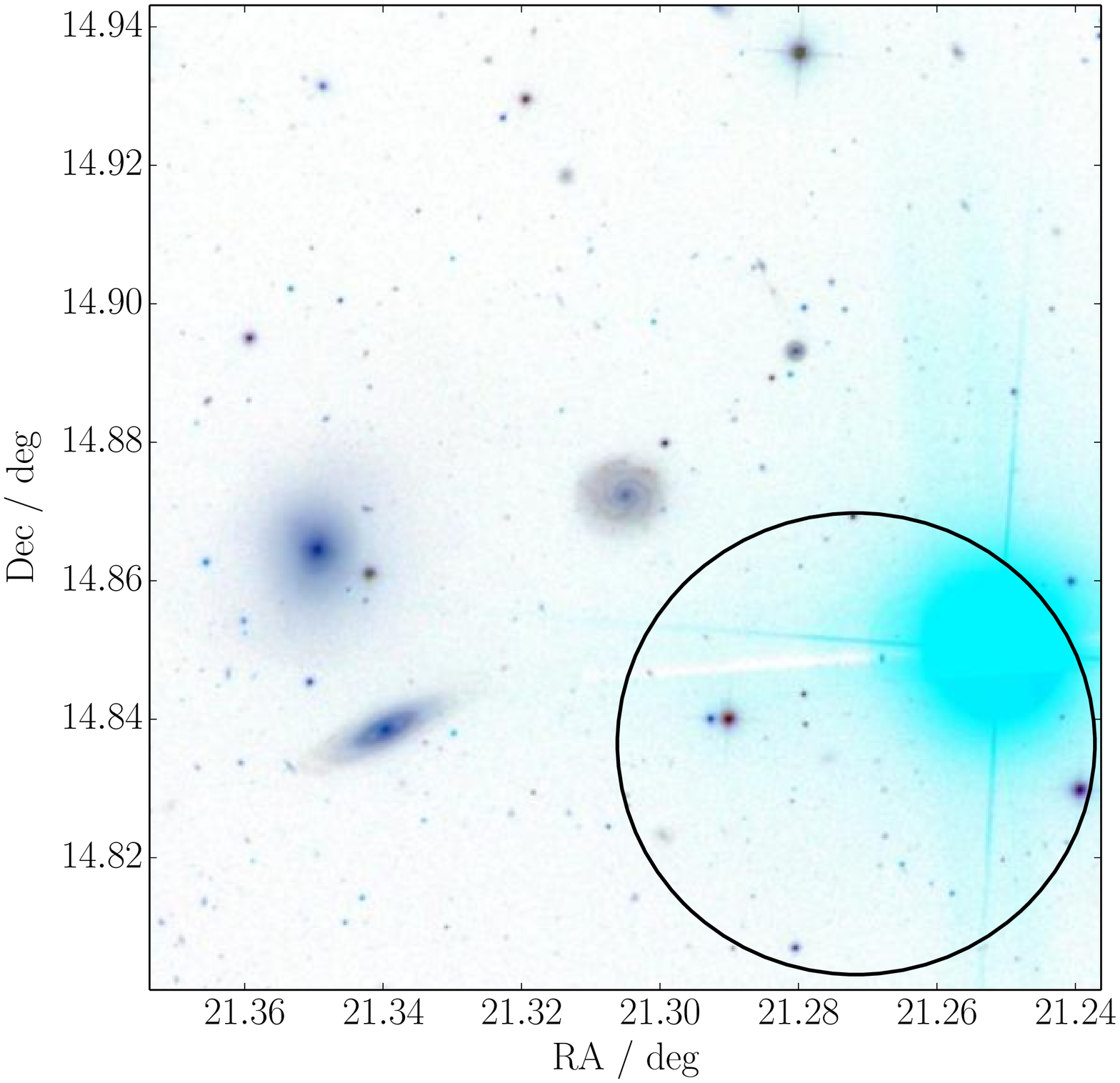}
\includegraphics[width=\columnwidth]{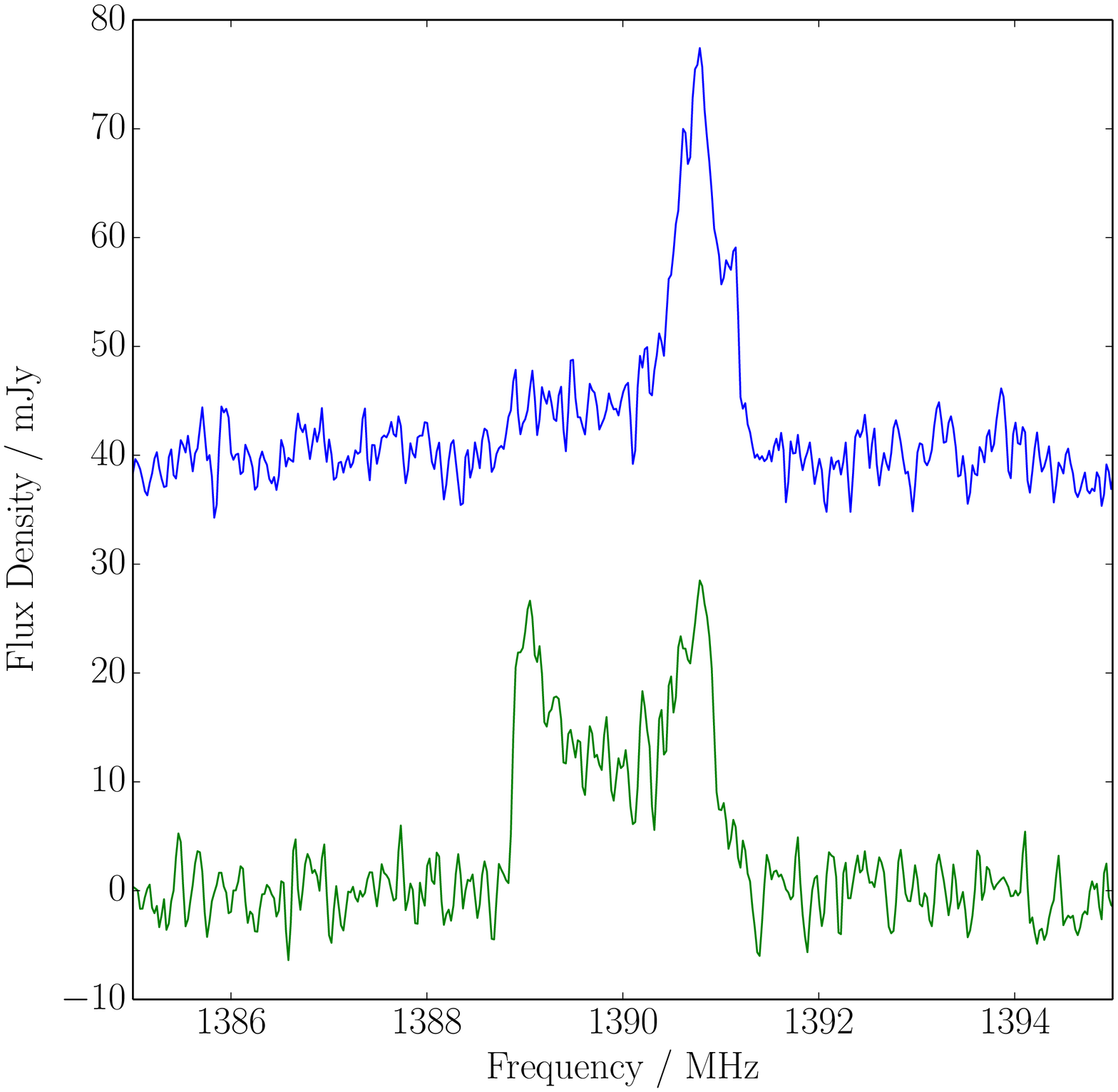}
\caption{The optical image (left) from the SDSS DR10 (http://skyserver.sdss3.org/dr10/en/tools/chart/image.aspx, \citet{Ahn+2014}) shows three galaxies; UGC978 and UGC983, with their respective ALFALFA spectra (right), and an early-type galaxy to the east, which ALFALFA does not detect. UGC978 is the central, face on spiral, its spectrum is the upper (blue), narrow profile, vertically offset by 40 mJy. UGC983 is the edge-on late-type galaxy to the south-east, associated with the lower (green), broad spectrum. The dark circle represents the ALFA beam on the sky (here taken to be a conservative 4'). Low levels of confusion are clear in the spectrum of UGC978, where there is excess emission over the velocity range of UGC983.}
\label{fig:conf_ex}
\end{figure*}

Source confusion is an issue for all galaxy surveys as blended sources lead to incorrect fluxes, masses, sizes, velocity widths and of course, number counts. In the submillimetre, source confusion is common as the surveys typically have poor resolution (compared to optical) and are at high redshift where source density is much higher; as a result submillimetre sources frequently overlap on the sky, often multiple times \citep[e.g.][]{Nguyen+2010}. In optical surveys, the high angular resolution and relatively low redshift (compared to submillimetre) makes confusion much less common, with it usually only occurring in the direction of clusters or in interacting systems (where the confusion is physical, not due to survey limitations).

If an optical survey had the resolution of a single dish HI survey, it would be impossible to pick out individual galaxies, every source would be confused, multiple times. It is only because HI astronomy is intrinsically spectroscopic that such 21cm surveys are possible, and confusion is actually uncommon. In this sense HI surveys present a unique variant of confusion. 

Unlike in the optical or submillimetre, galaxies are essentially transparent to 21cm radiation \citep[e.g.][]{Giovanelli+1994}. This means that 2-dimensional overlap on the plane of the sky is not a sufficient condition for sources to be confused. As well as overlap on the sky, the emission must overlap in redshift space. That is to say, that the sum of the observed velocity widths of the sources must be greater than twice their separation in redshift. As in most cases HI galaxies subtend an angle smaller than the telescope beam, a conservative condition for overlap on the sky would be if the two sources are within a beam diameter of each other. If both these conditions are met then the two sources will be confused to some degree.

Depending on the severity of the blend, confused sources may be extracted as single, or separate sources. However, in both cases this will introduce bias. When extracted as one source, that one source will have the flux (mass) of the combined sources, the velocity width may be increased, and the position of peak emission may be altered, potentially effecting the redshift and misleading the process of identifying a counterpart at other wavelengths. When extracted separately, all the same issues are possible to a lesser degree, as flux can bleed from one source to another. This also introduces an additional bias, as some flux (mass) is counted multiple times.

These biases can potentially influence the global data products of such surveys; correlation functions (CF), HI mass functions (HIMF), and HI velocity width functions (WF). While the CF will only be affected on small scales, the effect on the HIMF and WF is less straightforward. Furthermore, as the rate of confusion will depend on the physical size of the telescope beam at a given redshift, as well as the channel width, such biases will be dependent on redshift and survey instrument, likely leading to different surveys harbouring different biases in these functions used to describe and test cosmology and the growth of structure.

Recent works such as \citet{Moorman+2014}, \citet{Zwaan+2005}, and \citet{Springob+2005} have begun to look for environmental dependence of the HIMF. Such a dependence would be expected from a $\Lambda$CDM model of structure growth, as voids are expected to have an excess of low mass halos relative to filaments \citep[e.g.][]{Peebles2001}. However, confusion will likely also be influenced by environment, with more blends occurring in high density regions. It is necessary to have a more complete understanding of confusion in order to be sure any trends observed are really cosmological in origin.

With the commissioning of Square Kilometre Array precursors, many large area, blind surveys are expected. While there have been some estimates of confusion for these surveys \citep[e.g.][]{Duffy+2012b,Duffy+2012c}, it has primarily (as with current surveys) been ignored under the assumption that it will not have a significant impact. \citet{Duffy+2012b} (hereinafter \DMS) used a 1-dimensional CF and a fixed velocity range to make an estimate of the rate of confusion around the `knee'-mass of the HIMF, while \citet{Duffy+2012c} used semi-analytic models to populate halos from N-body simulations with HI gas, from which they derived an array of predictions for upcoming ASKAP (Australian Square Kilometre Array Pathfinder) surveys, including estimates of confusion. However, neither of these studies estimated the potential impact on the measurement of the HIMF.

Here we take an alternative approach, using both the 2-dimensional CF and mass-velocity width function (MWF) to derive an integral expression for the rate of confusion at a given distance, for any survey based on its resolution, depth and rms noise level. Present and future surveys are also simulated by drawing HI masses and velocity widths from the MWF, while neighbour separations are drawn from the 2D CF, allowing us to calculate the HIMF for confused and unconfused cases.

Our primary dataset, from which we derive the properties of our model, consists of the 40\% ($\alpha$.40) catalogue \citep{Haynes+2011} from the Arecibo Legacy Fast ALFA (Arecibo L-band Feed Array), or ALFALFA, survey \citep{Giovanelli+2005}, but we also make extensive use of the HI Parkes All Sky Survey \citep[HIPASS,][]{Barnes+2001}, to test our model and make comparisons. The ALFALFA survey, which has now completed data acquisition, covers approximately 6900 sq deg of sky, detects HI galaxies out to a redshift of 0.06, and was carried out using the 305m Arecibo telescope in Puerto Rico. Observations were completed in October 2012, with an average `open shutter' time efficiency of greater than 95\% including all startup, shutdown and calibration procedures. The ALFALFA team are currently reducing and extracting sources from the remaining dataset. HIPASS was carried out with the 64m Parkes telescope in New South Wales, and covers a greater area of sky than ALFALFA (approximately a hemisphere), but is less deep, detecting galaxies out to a redshift of 0.04. The $\alpha$.40 catalogue contains 11,941 high S/N extragalactic sources, almost all of which have optical counterparts identified in the Sloan Digital Sky Survey DR7 \citep{Abazajian+2009}, and the HIPASS catalogue HICAT \citep{Meyer+2004,Zwaan+2004} contains 4,315 sources.

The following section describes the model used to predict confusion rates for general surveys, and discusses how the relevant properties are determined from the $\alpha$.40 catalogue. In section 3 we display the results of our model, compare them to existing surveys, explore the effect confusion has on the HIMF, discuss predictions for proposed upcoming surveys, and evaluate the limitations confusion places on single dish telescopes. Finally, section 4 outlines our conclusions and recommendations for dealing with confusion.

\section{Modelling Confusion}
\label{sec:model}

Unlike optical or submillimetre surveys, in HI radio surveys confusion must be spectroscopic; it requires overlap both on the plane of the sky and in velocity space. An example of two confused sources from the $\alpha$.40 catalogue is shown in figure \ref{fig:conf_ex}. The galaxy in the centre of the frame is a face on spiral galaxy (UGC978), with a narrow profile (due to the projection), however there is a clear excess contribution (at lower frequency than the main peak of emission) that is coincident in frequency with the profile of another nearby galaxy (UGC983) within the beam. If the two galaxies were separated by an angular distance greater than the diameter of the beam, they would not be confused as flux could not be simultaneously received from both sources (ignoring the possibility of flux entering from spatial sidelobes). They would also not be confused if their redshifts were different by an amount larger than half the sum of their velocity widths, as then their emission would not be overlapping in frequency. This would still be true even if they were in contact on the plane of the sky.

The model of confusion will be explained by beginning with an idealised case and replacing each component until a realistic model is reached. The details of the fits used to describe the correlation function, mass-width function and the detection limit can be found in the appendix. 

To model how frequently this kind of dual overlap occurs, consider a Universe where all galaxies are the same mass ($M_{0}$), with the same projected velocity width ($W_{0}$), and are distributed randomly in 3D space with a mean number density $n_{0}$. In order for two galaxies to be blended in a survey they would need to be both closer together in projected linear distance ($\kappa$) than the linear diameter of the beam at the distance to the galaxies, $D_{\mathrm{beam}}(d)$, and closer together along the line-of-sight ($\beta$) than the effective radial separation, $W_{0}/H_{0}$ (where $H_{0}$ is the Hubble constant, $\sim70 \, \mathrm{km\,s^{-1}\,Mpc^{-1}}$). 

The diameter of the telescope beam, rather than its radius, is used because the surveys considered are blind, meaning that in general a source can be anywhere within the beam, and so other emission from anywhere within a beam's width of that source could potentially contribute to its measured flux. The maximum line-of-sight separation, $W_{0}/H_{0}$, results from the requirement that the velocity (or equivalently, redshift) difference between the two sources must be less than half the sum of their velocity widths, in order for their velocity profiles to overlap (see figure \ref{fig:conf_ex}). Thus, the criteria for two sources to be confused are:
\begin{equation}
\centering
\label{eq:blend_crit_a}
\kappa < \Theta_{\mathrm{beam}} d 
\end{equation}
\begin{equation}
\centering
\label{eq:blend_crit_v}
-\frac{W_{0}}{H_{0}} < \beta < \frac{W_{0}}{H_{0}}
\end{equation}
where $\Theta_{\mathrm{beam}}$ is the angular diameter of the telescope beam, and $d$ is the comoving distance to the central source. Here the phrase ``central source" refers to the source at the centre of the cylindrical volume being considered, this does not necessarily imply that it was at the centre of the beam when detected.

According to the Poisson distribution, the probability of a blend occurring (i.e. one or more galaxies lying in the cylindrical volume defined by equations \ref{eq:blend_crit_a} \& \ref{eq:blend_crit_v}) is:
\begin{equation}
\centering
P(\mathrm{blend}) = 1 - e^{-\left< N \right>},
\end{equation}
where $\left< N \right>$ is the average number of additional sources expected to be found within the relevant cylindrical volume around the central source. In this model $\left< N \right>$ can be found simply by multiplying the number density of sources by the cylindrical volume:
\begin{equation}
\centering
\left< N \right> = 2 \pi n_{0} \Theta_{\mathrm{beam}}^{2} d^{2} \frac{W_{0}}{H_{0}}.
\end{equation}
Within this uniform model $\left< N \right>$ grows quadratically with distance, as the volume increases with the square of the physical size of the beam.
\begin{figure*}
\centering
\includegraphics[width=\textwidth]{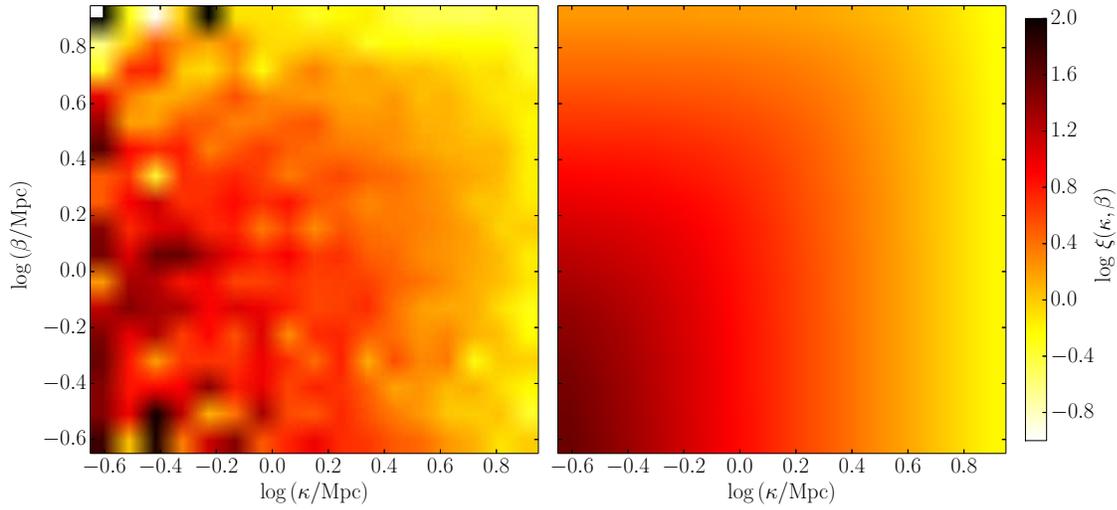}
\caption{The 2-dimensional correlation function of the ALFALFA 40\% sample (left), calculated by \citet{Papastergis+2013}, and our fit using an elliptical shaped function in the projected separation - line-of-sight velocity ($\kappa$-$\beta$) plane (right). The slight elongation in the velocity direction indicates a weak `finger of god' effect.}
\label{fig:2DCF}
\end{figure*}

\begin{figure*}
\centering
\includegraphics[width=1.5\columnwidth]{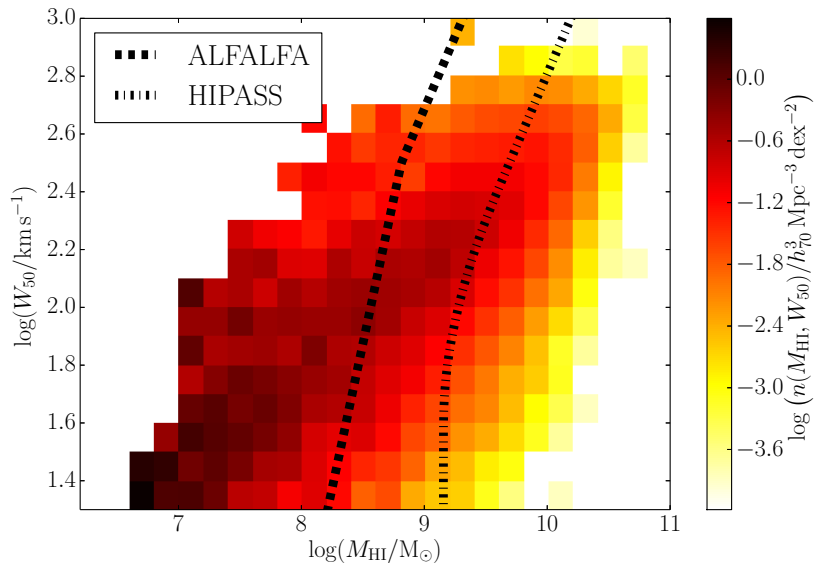}
\caption{The ALFALFA mass-width function (see appendix of \citet{Papastergis+2014}). Each pixel represents the intrinsic number density of HI galaxies with those mass and velocity width properties. The HIMF is the integral through all velocities, and the mass conditional velocity width function (MCWF), is a vertical slice at the relevant mass. The ALFALFA and HIPASS 50\% completeness limits at 50 Mpc are shown as dashed and dot-dash lines, respectively \citep{Haynes+2011, Zwaan+2004}. Integrals over all detectable sources cover all the space to the right of these lines.}
\label{fig:detlim}
\end{figure*}

This is the most basic model of spectroscopic confusion, and in order to construct a more comprehensive model each component must be realistically accounted for. Firstly, to address the fact that the Universe is not uniform on the scale of galaxy neighbour separations we must employ the correlation function (CF), the excess probability (above random) of two galaxies being separated by a given distance. \citet{Papastergis+2013} measured the CF of the $\alpha$.40 dataset, which is plotted in figure \ref{fig:2DCF} along with our 2D fit. The $\kappa$-direction corresponds to linear separations perpendicular to the line-of-sight, and the $\beta$-direction corresponds to separations along the line-of-sight, both are measured in Mpc. 

The inclusion of the CF, $\xi(\kappa,\beta)$, alters the calculation of the occurrence rate, $\left< N \right>$. When evaluating the integral over the volume defined by the beam and maximum possible line-of-sight separation given the velocity widths, the probability that a galaxy will be found at any given point is now multiplied by $1+\xi(\kappa,\beta)$. This gives the occurrence rate as
\begin{eqnarray}
\label{eq:occ_rate_simp}
\left< N \right> = 2 n_{0} \int_{0}^{\frac{W_{0}}{H_{0}}} \int_{0}^{\Theta_{\mathrm{beam}} d} 2\pi \kappa \left( 1+\xi(\kappa,\beta) \right) \, \mathrm{d}\kappa \, \mathrm{d}\beta.
\end{eqnarray}

Next, consider galaxy masses and velocity widths. Rather than fixed values they should be drawn from distributions representative of the intrinsic properties of HI galaxies. For masses this distribution is the HIMF ($\phi(M)$), and for velocity widths it is the WF. However, since the two properties are not independent the mass conditional velocity width function (normalised such that it integrates to unity over all widths) $p(W|M)$, is the appropriate distribution to use. We use the ALFALFA HIMF as calculated by \citet{Martin+2010}, and follow a similar procedure to the appendix of that paper to calculate the mass conditional width function (MCWF), the details of which can be found in the appendix. 

The ALFALFA mass-width function (MWF) is shown in figure \ref{fig:detlim}. The HIMF is this function integrated through all possible velocity widths, whereas the MCWF can be thought of as a slice through all velocities, at a particular mass. The ALFALFA 50\% completeness limit \citep{Haynes+2011} at a particular distance (50 Mpc) is shown as the dashed black line, and the equivalent limit for HIPASS is the dash-dot black line \citet{Zwaan+2004}. When integrating over all detectable masses and velocity widths, as we will do below, the integral simply covers everything to the right and below the appropriate line.

Now that there are a range of possible masses and velocity widths for the second galaxy, instead of multiplying by $n_{0}$ in the expression for $\left< N \right>$, all possible masses and widths, weighted by the probability of them occurring, must be integrated through. Thus, the occurrence rate now becomes
\begin{eqnarray}
\label{eq:occ_rate}
\left< N \right> &=& 2 \int_{W_{\mathrm{min}}}^{W_{\mathrm{max}}} \int_{M_{\mathrm{lim}(d,W_{2})}}^{M_{\mathrm{max}}} \phi(M_{2}) p(W_{2}|M_{2}) \nonumber\\
&& \int_{0}^{\frac{W_{1}+W_{2}}{2 H_{0}}} \int_{0}^{\Theta_{\mathrm{beam}} d} 2\pi \kappa \left( 1+\xi(\kappa,\beta) \right) \nonumber\\
&&\, \mathrm{d}\kappa \, \mathrm{d}\beta \, \mathrm{d}M_{2} \, \mathrm{d}W_{2},
\end{eqnarray}
where $W_{1}$ and $W_{2}$ are the velocity widths of the central galaxy and the galaxy it is potentially blended with, and $M_{2}$ is the HI mass of this second galaxy. $W_{\mathrm{min}}$ and $W_{\mathrm{max}}$ are the limiting velocity widths, taken to be 15 and 1000 $\mathrm{km\,s^{-1}}$ respectively, $M_{\mathrm{max}}$ is the maximum HI mass considered ($10^{11} \mathrm{M_{\odot}}$), and $M_{\mathrm{min}}(d,W)$ is the minimum detectable mass for a given velocity width, at a given distance (although an absolute minimum is set at $10^{6.2} \mathrm{M_{\odot}}$). As before, $\left< N \right>$ can be used to estimate the probability of a blend: $P(\mathrm{blend}) = 1 - e^{-\left< N \right>}$. 

Implementing realistic values of mass, velocity width, and the detection limit have two important effects. The line-of-sight separation that can result in confusion will now be dependent on the velocity widths of each pair of galaxies that might be confused. Thus, similarly to equation \ref{eq:occ_rate} we must integrate through all possible masses and widths for the central galaxy, with each mass and width weighted appropriately, and again truncating the integral at the detection limit. This gives our final model as:
\begin{eqnarray}
\label{eqn:blendprob_simp}
&&P(\mathrm{blend}|d) = \frac{1}{n_{\mathrm{Det}}(d)} \int_{W_{\mathrm{min}}}^{W_{\mathrm{max}}} \int_{M_{\mathrm{lim}}(d,W_{1})}^{M_{\mathrm{max}}} \nonumber \\
&&\phi(M_{1}) \, p(W_{1}|M_{1}) \left[ 1 - e^{-\left< N(d,W_{1}) \right>} \right] \; \mathrm{d}M_{1} \, \mathrm{d}W_{1}.
\end{eqnarray}
Here the normalisation, $n_{\mathrm{Det}}(d)$, is the number density of detectable sources at a given comoving distance, $d$. This is calculated by integrating the MWF over the detectable region of HI mass and velocity width (see figure \ref{fig:detlim}).

The above equation represents the specific case of confusion between detectable sources only, which will be the blends that are noticeable in the final dataset of a survey. However, sources may also be blended with objects that are below the detection limit. To assess how frequently such blends occur the exact same framework can be used, but instead of setting the lower bound of the integration over $M_{2}$ (in the expression for $\left< N \right>$) by the detection threshold, it should be set as the minimum mass object considered as a source of confusion. In section \ref{sec:fut_survey} we consider various different prescriptions for what minimum mass object constitutes a significant source of confusion.

Although this model now encompasses realistic masses and velocity widths, as well as the distribution of sources on the sky and in redshift space, it still assumes (as in equations \ref{eq:blend_crit_a} \& \ref{eq:blend_crit_v}) that both the beam response and the velocity profiles of galaxies are top-hat functions, clearly this is a crude simplification. However, as we show in the following section, this simple model reproduces the observed rate of confusion in both ALFALFA and HIPASS, and can be used to make an estimate of the upper limit of the impact this has on the shape of the HIMF.

In general this model cannot be evaluated analytically, and so we carry out a Monte Carlo integration to estimate the rate of confusion as a function of redshift. While the data itself could be used to describe the HIMF, MCWF and 2D CF, we instead make analytic fits to each of these (described in detail in the appendix) in order to produce a more accessible model and to reduce computation time.

\subsection{Catalogue Simulation}
\label{sec:mod_sim}

In order to evaluate the impact confusion has on the HIMF, it is necessary to explicitly simulate a catalogue of blended and non-blended HI detections, so that an HIMF can be derived for both cases.

The survey volumes were simulated by drawing masses and widths from the HIMF and MCWF (described in the appendix), placing them randomly in space with the average number density associated with the ALFALFA HIMF, and then eliminating anything below the detection limit of the relevant survey. 

Confusion was assessed for each source by drawing the number of neighbours within 1000 km/s and the beam width from the expression for $\left< N \right>$ (equation \ref{eq:occ_rate}), and then assigning their positions (relative to the central) galaxy by drawing from the 2D CF in the same range. Masses and widths were then drawn as for any other galaxy (but all were retained, even those below the detection limit), at which point it can be assessed whether they are blended with the central galaxy.

\section{Results \& Discussion}

This comprehensive model of confusion must now be tested against existing blind HI surveys. Good agreement with ALFALFA and HIPASS is demonstrated before this model is used to make predictions for upcoming surveys.

\subsection{Existing Surveys: Rate of Confusion}

To test the validity of the model described in section 2, we wish to compare its results to those of existing blind HI surveys, in this case HIPASS \citep{Meyer+2004} and ALFALFA's 40\% sample, $\alpha$.40 \citep{Haynes+2011}.

Both surveys are modelled based on their published detection limits. For ALFALFA this corresponds to setting a sharp cut off at 50\% completeness (as defined in \citet{Haynes+2011}, however the HIPASS completeness surface is more complicated \citep{Zwaan+2004}, being a function of both peak and integrated flux. Thus, HIPASS is only simulated directly (as described in section \ref{sec:mod_sim}), rather than run through our integral models. The detection limit used here is cut at 50\% completeness, and above that the completeness function of each source is treated as a probability of detection. Here we note that this formulation, based on the ALFALFA MWF and the published completeness limits, produces appropriate number counts, HI mass and velocity width distributions for both ALFALFA and HIPASS, despite the fact that the published HIMFs of the two surveys are different.

\begin{figure*}
\includegraphics[width=\columnwidth]{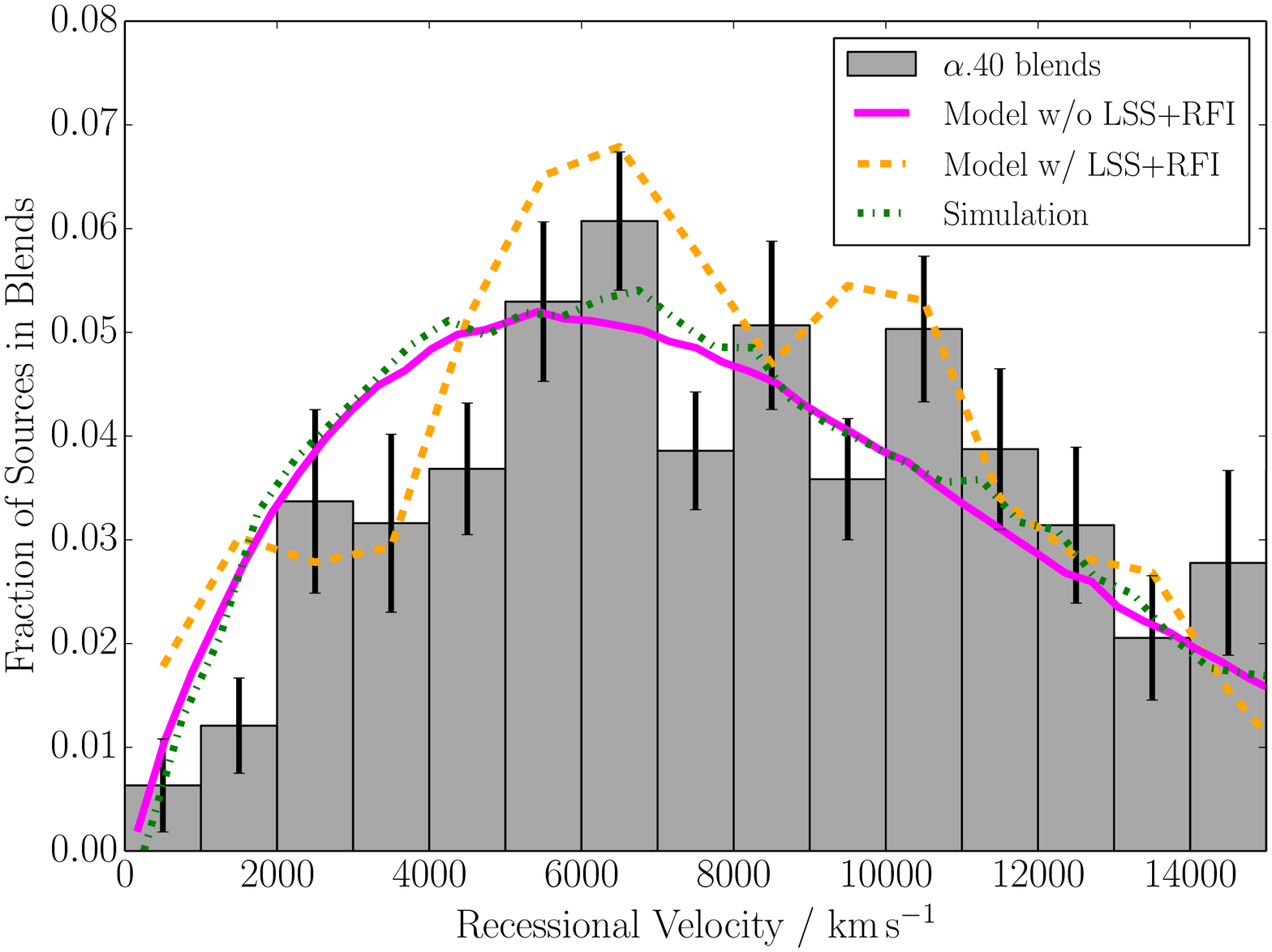}
\includegraphics[width=\columnwidth]{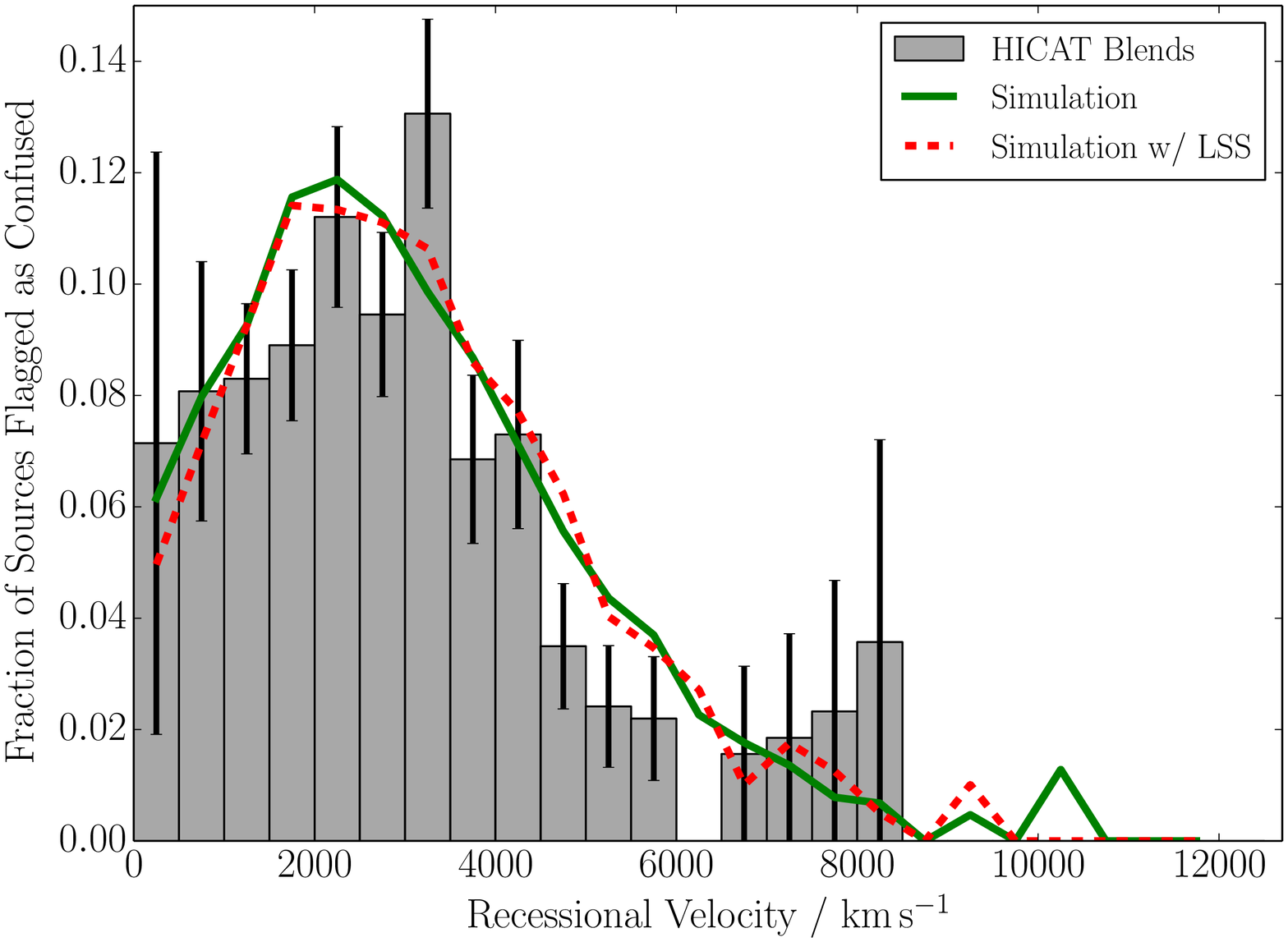}
\caption{The observed rates of blended sources in ALFALFA (left) and HIPASS (right), compared to the model of confusion between detectable sources only (solid magenta), and confusion in a simulated population (dot dashed green). The ALFALFA data is binned in bins that are 1000 $\mathrm{km\,s^{-1}}$ wide, and is cut off at 15,000 $\mathrm{km\,s^{-1}}$, beyond which a significant band of RFI makes the completeness of the survey difficult to model. The HIPASS blends are binned in 500 $\mathrm{km\,s^{-1}}$ wide bins. The fit to ALFALFA is improved by including weighting for LSS (from 2MRS) and RFI (dashed orange), though there are still discrepancies which are discussed in the text. The LSS correction has little impact in the case of HIPASS (dashed red), indicating that it was a small bias to begin with. The plotted error bars include only counting errors. ALFALFA detects a number of blends between nearby galaxies and tidal debris, we make no attempt to model these complex systems, and such sources are not included here.}
\label{fig:AH_conf}
\end{figure*}

\begin{figure*}
\includegraphics[width=\columnwidth]{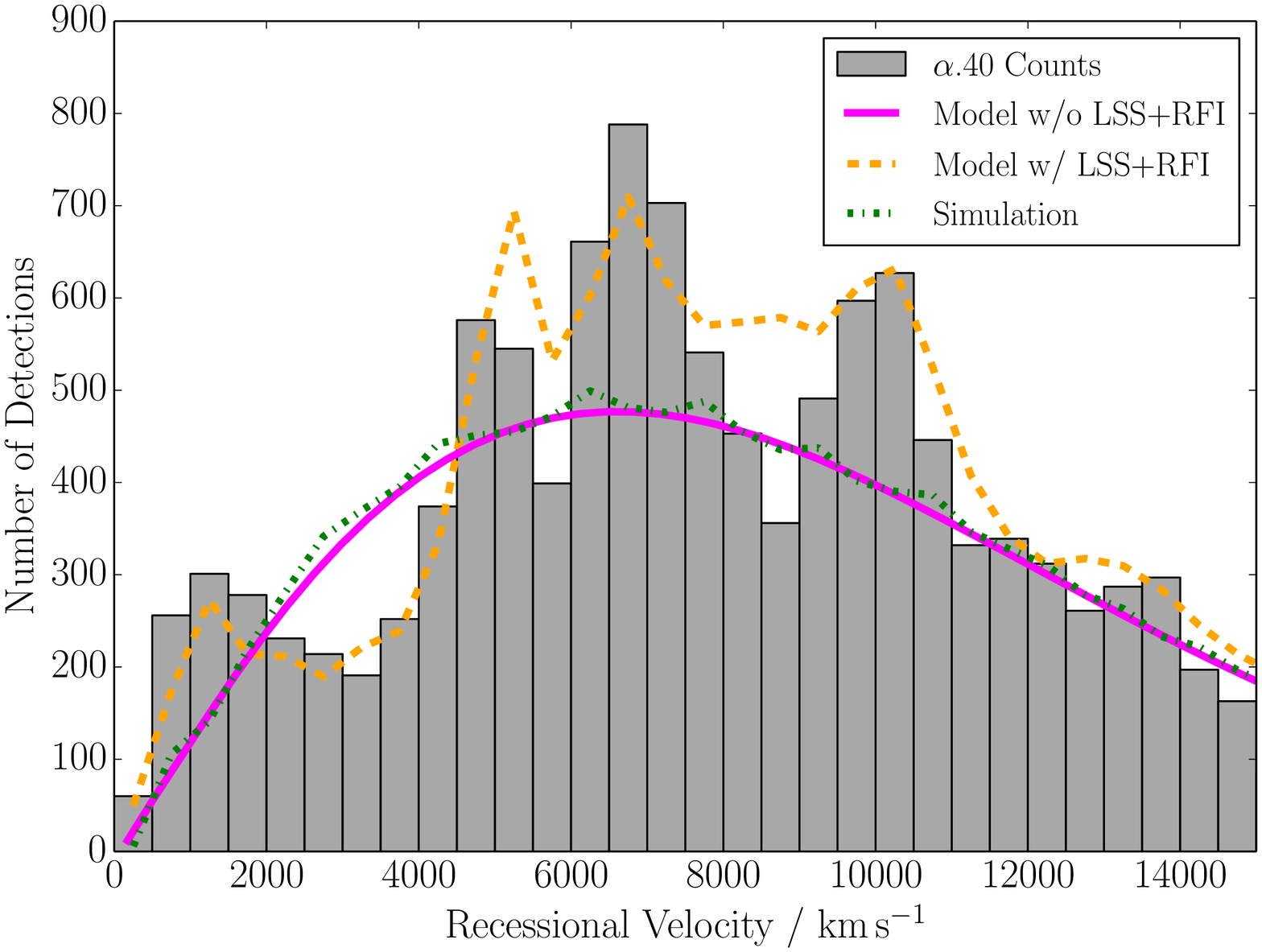}
\includegraphics[width=\columnwidth]{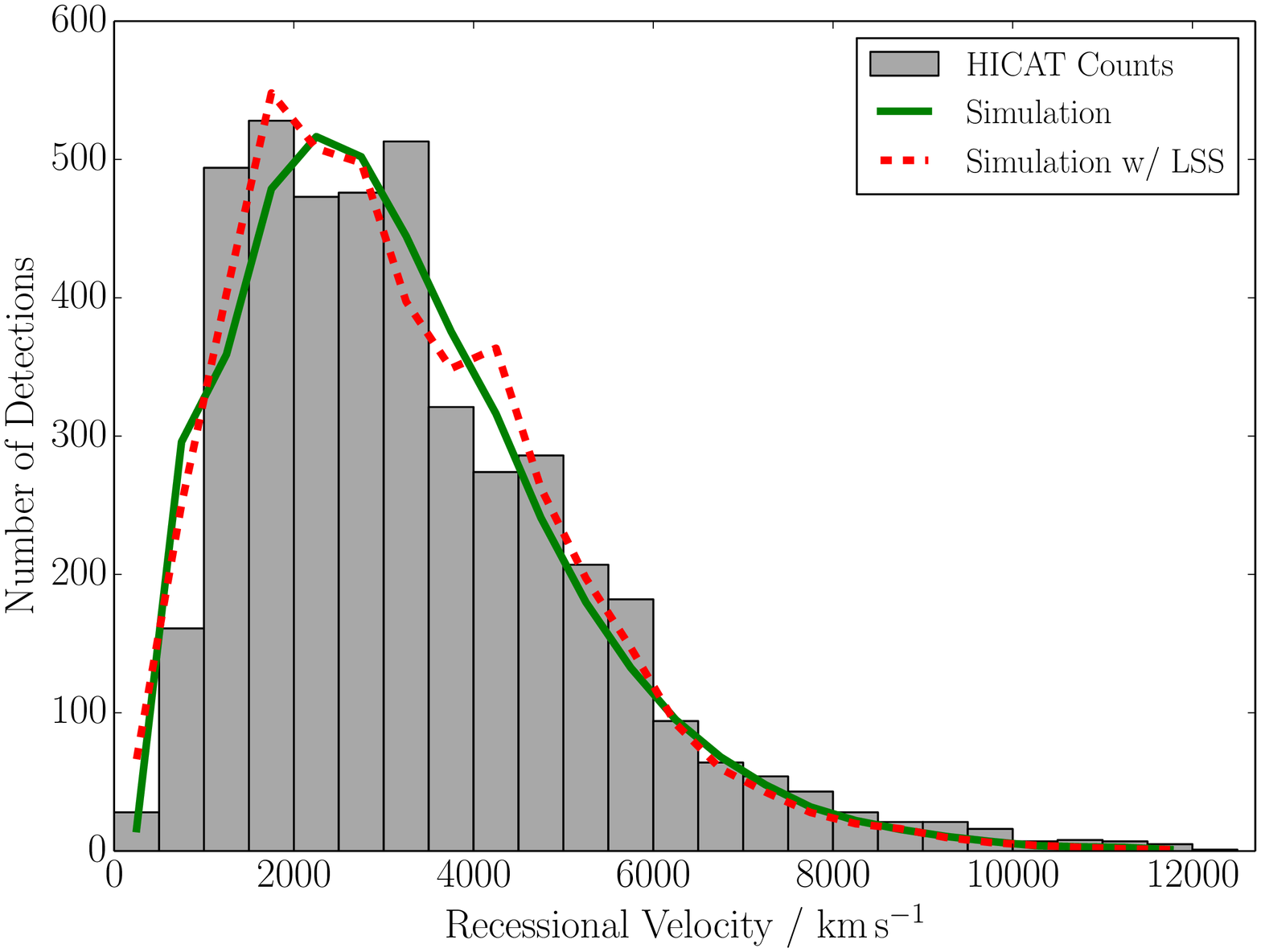}
\caption{The observed detection number counts in 500 $\mathrm{km\,s^{-1}}$ wide bins for ALFALFA (left) and HIPASS (right). The solid magenta line shows the equivalent number counts from our model and the green lines show the number counts from simulations, both without corrections for LSS or RFI, and the dashed orange line shows the model with those corrections for ALFALFA, while the dashed red line shows the HIPASS simulation with the LSS correction.}
\label{fig:AH_num}
\end{figure*}

In order to make a fair comparison with the data, the occurrence rate of blends between detectable sources only, was calculated. The equivalent value for the real data sets can be measured by counting the number of sources that are within a beam's width of another detected source, and within half the sum of their velocity widths of each other in velocity space. We carry out this measurement for the $\alpha$.40 catalogue, and use an equivalent flag set in the HIPASS source catalogue (HICAT). The estimated rates from the surveys are shown as the bars in figure \ref{fig:AH_conf}, the model is the magenta line, and the green line represents the simulated catalogue. The same colour scheme is used in figure \ref{fig:AH_num} to show the observed and modelled number counts as a function of redshift.

It can be seen that the models are reasonable fits to $\alpha$.40 and HIPASS confusion rates, though the deviations are larger for the $\alpha$.40 volume. The reason for this discrepancy is that $\alpha$.40 contains significant background density variations due to large scale structure (and radio frequency interference), whereas the larger sky area of HIPASS effectively averages out this bias. The ALFALFA confusion rate is plotted in wider bins in order to smooth the effects of large scale structure (LSS), but in addition we also account for LSS by weighting the background density of HI sources using a full sky 3D overdensity map from the 2MASS Redshift Survey (2MRS), calculated by \citet{Erdogdu+2006} (provided by P. Erdo{\v g}du and C. Springob via private communication). The fraction of the survey volume eliminated by radio frequency interference (RFI) as a function of redshift was calculated in \citet{Papastergis+2013} (their figure 6), and in addition to weighting by LSS we also weight the intrinsic number density by the fraction of the volume available in the presence of RFI.

The ALFALFA-like model with weighting for LSS and RFI now fits somewhat better (see figures \ref{fig:AH_conf} \& \ref{fig:AH_num}), but there are still a few discrepancies. The largest of these discrepancies occurs at approximately 1,500 $\mathrm{km\,s^{-1}}$, where there is an over prediction of blends in the model. This can be explained by the presence of the Virgo cluster. While this represents a significant overdensity, leading to an excess of detections, it does not produce the corresponding excess of blends. Given Virgo's proximity it is possible to detect galaxies in HI much closer to the centre of the cluster than with any other cluster, which leads to very large peculiar velocities, making confusion less likely than predicted by a model without this level of complexity. In addition, galaxies in Virgo are HI-deficient \citep{Solanes+2002} which could decrease their observed HI velocity widths, also reducing the chance of confusion.

In addition to reproducing the observed rate of blends between detections, it is also important to check that the model and simulations can reproduce the observed detection counts of the surveys, as a function of redshift. Not only is this a critical criteria for accurately modelling a survey, it is also one of the most important quantities in determining the blend rate. As can be seen in figure \ref{fig:AH_num}, both the number counts of ALFALFA and HIPASS are approximately reproduced, though ALFALFA requires a LSS and RFI correction to achieve a convincing match.

\begin{figure}
\includegraphics[width=\columnwidth]{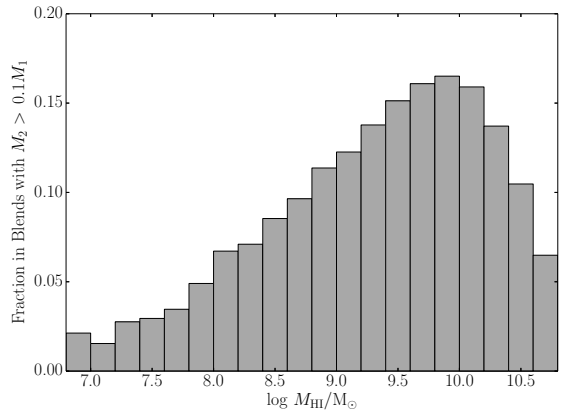}
\caption{The fraction of simulated ALFALFA detections in blends with other galaxies above 10\% of their own HI mass (regardless of detectability), in logarithmic bins of width 0.2 dex. The peak rate of confusion occurs around the `knee'-mass of the HIMF. Below this mass the velocity widths of galaxies drop, making blending less likely, and above this mass the number density of sources with appropriate masses drops exponentially with mass.}
\label{fig:conf_m}
\end{figure}

As objects are often studied in classes defined by mass (for example dwarfs, or $M_{*}$ galaxies), an understanding of the relative rates of confusion across such classes is of interest. Figure \ref{fig:conf_m} shows the rate of confusion of simulated ALFALFA sources with another galaxy at least 10\% of the central's HI mass, binned by mass. This represents only the blends where there is the potential for a non-negligible alteration of the observed mass. The highest rate of confusion occurs around the `knee' of the HIMF function, with it dropping off approximately exponentially in either direction in mass. Essentially identical behaviour was seen in all our simulations, only the amplitude varied from survey to survey.

The above behaviour can be understood as follows: occurrences of confusion will become more likely as mass increases, because galaxy velocity widths grow with mass. This greatly increases the cylindrical volume available to confusion, as an increase in velocity width of 70 $\mathrm{km\,s^{-1}}$, increases the depth of the cylinder by approximately 2 Mpc, whereas typical angular scales will correspond to tens or hundreds of kpc. In addition, the `finger of god' effect causes there to be more power in the line-of-sight direction (compared to the perpendicular direction), than would be expected from a model using a 1D CF. However, once beyond the `knee' of the mass function the availability of other sources of comparable mass, drops precipitously, and the increase in velocity width begins to stagnate, leading to a decline in the occurrence of these blends in the most massive sources.

\subsection{Existing Surveys: Bias in the HI Mass Function}

Figures \ref{fig:AH_conf} \& \ref{fig:AH_num} give a strong indication that this model is valid, as it is able to simultaneously reproduce the detection rate of both surveys with redshift, and the observed rate of confusion. However, as well as knowing how much confusion is present in a survey it is important to understand what effect this has on the measured quantities, such as the HIMF. To do this we make use of the simulated catalogues of each survey (see section \ref{sec:mod_sim}).

The HIMF is calculated using the $1/V_\mathrm{max}$ method (as there is no LSS included). For the non-confused HIMF only detectable galaxies are considered, but for the confused HIMF all sources confused with their central (detectable) object are considered together as a single source. The exact details of how the flux and velocity width are affected in a blend will depend strongly on the separation and geometry. Here we aim to estimate upper limits on the influence of confusion, so we make the extreme assumption that the velocity width of the central source is unchanged, but the flux (mass) becomes the sum of all objects that are blended together.

\begin{figure*}
\includegraphics[width=\columnwidth]{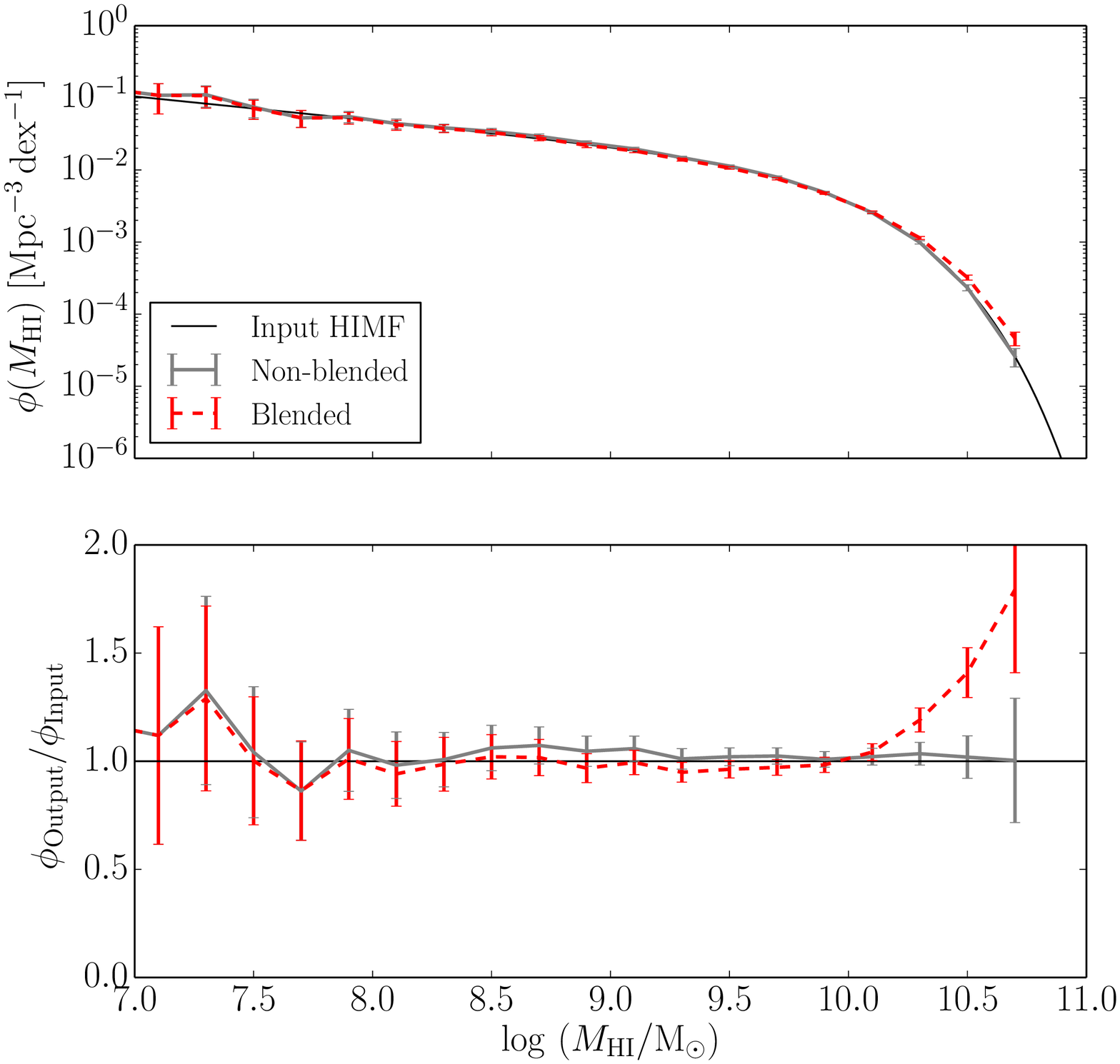}
\includegraphics[width=\columnwidth]{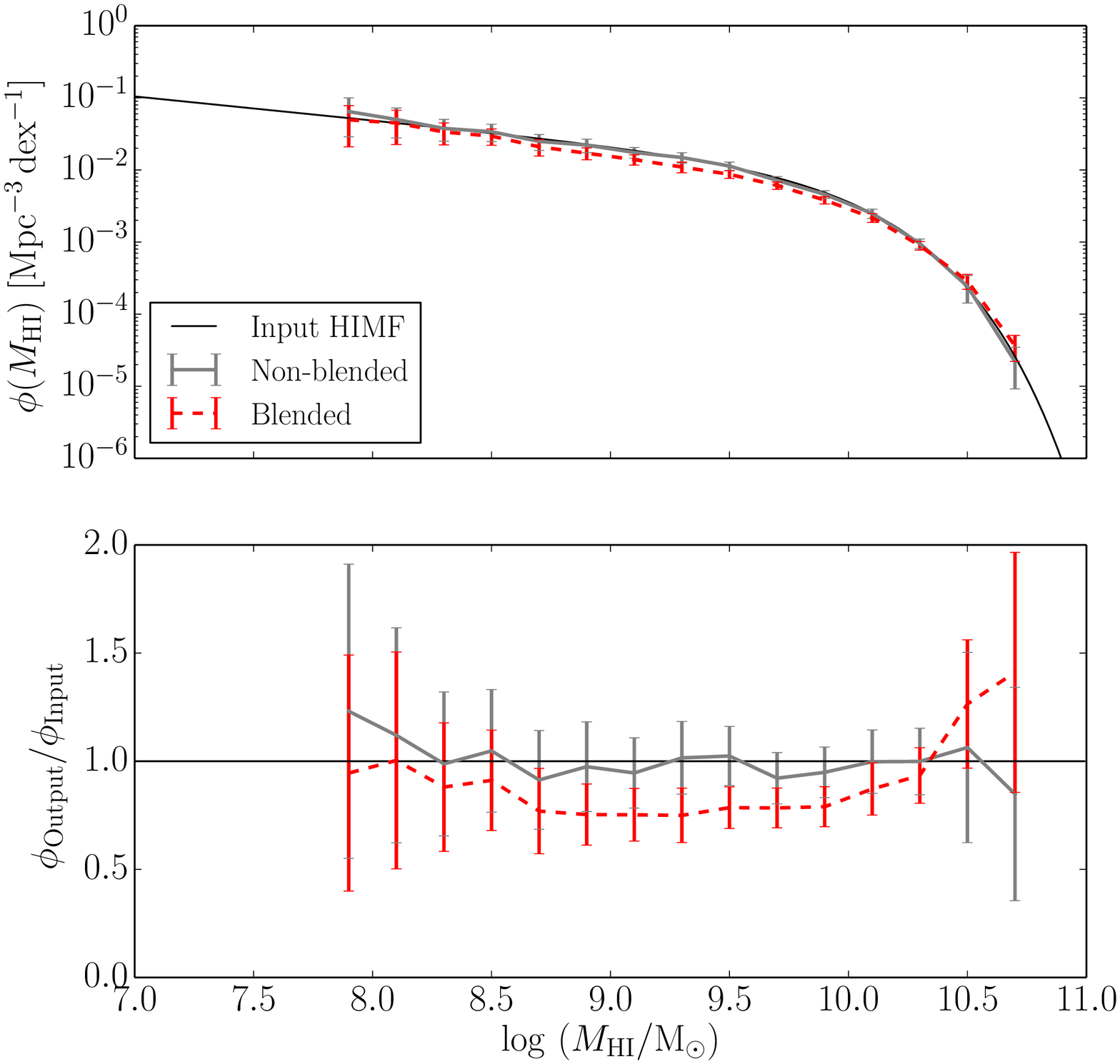}
\caption{Example HIMFs (top row) for simulated ALFALFA (left) and HIPASS (right) surveys, and their fractional deviations from the simulation's input HIMF (bottom row). The thin black line represents the input HIMF, the thick grey line is the calculated HIMF in the absence of confusion, and the dashed red line is the HIMF with confusion. The error bars are errors purely from counting noise, as these simulations contain no LSS or RFI. The effect of confusion is to depress the faint end slope and enhance the values beyond the `knee'. This results in measuring a marginally steeper faint end slope and a greater `knee' mass, in the case where confusion is present.}
\label{fig:AH_HIMF}
\end{figure*}

Figure \ref{fig:AH_HIMF} shows the simulated HIMF, the solid grey lines are non-confused, and the dashed red are confused. The general action of confusion is to increase the mass of a given object, and potentially push it in to a higher mass bin. Its higher apparent mass fools you in to thinking it is detectable over a larger volume than it is. Therefore the overall influence on the shape is to decrease the HIMF in the original bin and enhance it in the apparent bin. The most noticeable effect occurs around and beyond the `knee', where the net result of these competing effects switches from the former to the latter. Galaxies just below $M_*$ can become blended together, causing the HIMF to be suppressed immediately before the `knee', and enhanced immediately after it, where the more massive, blended sources now fall and true sources become scarce. 

The alterations to the HIMF's shape can be measured by the deviations in the parameters of Schechter function fits. The faint end slope, $\alpha$, shows a slight decrease of less than 2$\sigma$ (compared to published random errors for the ALFALFA and HIPASS HIMFs) in both the simulations. For ALFALFA this decrease was 0.03, and for HIPASS it was 0.04, which corresponds to a 1-2$\sigma$ deviation in both cases. However, there was large variance between the values calculated in the 20 HIPASS simulations, whereas the 20 ALFALFA simulations were very consistent. The estimate of a decrease of 0.04 in faint end slope of HIPASS' HIMF corresponds approximately to the scale of the systematic error estimated by \citet{Zwaan+2004}. $M_{*}$, the `knee' mass, was more severely altered, showing a 2-3$\sigma$ increase, or 0.06 dex for both ALFALFA and HIPASS. In this case the alteration was more than double the previously estimated systematic error in the HIPASS HIMF. 

As the parameters of the Schechter function are highly covariant we also estimated the alteration to the faint end slope by fitting a straight line (in log-log space) to all mass bins below $10^{9} \mathrm{M_{\odot}}$. Though these results were significantly more noisy, the mean values were similar to those quoted above, giving decreases of 0.015 and 0.06 for ALFALFA and HIPASS, respectively.

The larger beam of the Parkes telescope compared to the Arecibo observatory leads one to expect that HIPASS would suffer much greater adverse effects of confusion, however the impact on the HIMF has many competing factors and is a non-linear function of the rate of confusion. The alteration of the faint end slope depends on relative, rather than absolute confusion, that is, the slope is dependent on the relative amount of confusion in adjacent bins. In other words, a more confused survey does not necessarily have a more altered faint end slope, so long as the suppression is nearly uniform along it. The `knee' mass is more simply related to the rate of confusion; it will always increase with increasing confusion (assuming the survey is not artificially truncated in redshift extent, see section \ref{sec:dish_lim}). The reason that HIPASS' $M_{*}$ is not significantly more impacted than it is for ALFALFA, is likely due to there being similar rates of total confusion (not just with other detections) at the respective distances where most of their $M_{*}$ galaxies are detected ($\sim 50$ and $\sim 150$ Mpc).

The overall effects of confusion are to slightly steepen the faint end slope ($\alpha$), though this is a weak effect, and increase the value of $M_*$, the position of the `knee'. This means that ALFALFA's 0.1 dex higher $M_*$ value, compared to the more confused HIPASS \citep{Martin+2010, Zwaan+2005}, cannot be explained by confusion. However, given the variance in the HIPASS simulations, its  steeper faint end slope could be a result of increased confusion. 

At this stage the reader should recall that these estimates are intended to be conservative, in that they aim to estimate the worst case scenario. Implementing realistic source angular sizes and velocity profiles, along with the beam response function would likely reduce the impact on the shape of the HIMF. Additionally, careful source extraction probably mitigates some of the biases caused by confusion.

Finally, an encouraging point is the relative insensitivity of the faint end slope to spectroscopic confusion. Although studies looking for environmental dependence of the HIMF \citep{Springob+2005, Zwaan+2005, Moorman+2014} are likely to include biases in the faint end slopes they derive, due to differing levels of confusion intrinsic to the regions being compared, a detection of a 3$\sigma$ deviation from ALFALFA's faint end slope would still be robust against the effects of confusion. However, caution should be used when comparing $M_*$ in different environments, as this is more noticeably biased by confusion.

\subsection{Predictions for Future Surveys}
\label{sec:fut_survey}

A number of blind HI galaxy surveys have been proposed recently, primarily as part of Square Kilometre Array (SKA) precursors, these include medium-depth surveys out to a redshift of about 0.25, and very deep surveys aiming to detect HI at redshifts of order unity. The Australian SKA Pathfinder telescope (ASKAP) plans to undergo two medium depth surveys, the Widefield ASKAP L-band Legacy All-sky Blind surveY (WALLABY - PIs: B. Koribalski \& L. Staveley-Smith) and the Deep Investigation of Neutral Gas Origins (DINGO - PI: M. Meyer), whist the Westerbork Synthesis Radio Telescope (WSRT) intends to carry out its own survey similar to WALLABY, but in the northern hemisphere, called the Westerbork Northern Sky HI Survey (WNSHS - PI: G. J{\' o}zsa). The deep surveys are COSMOS (Cosmological Evolution Survey) HI Large Extragalactic Survey (CHILES - PI: J. van Gorkom), currently underway at the Very Large Array (VLA), and the proposed Looking At the Distance Universe with MeerKAT survey (LADUMA - PIs: S. Blyth, B. Holwerda \& A. Baker). In this section we ask how confused these next generation, deeper survey will be, and how this will affect their ability to measure the HIMF and its evolution with redshift.

\citet{Duffy+2012c} published predictions of the rms noise and channel widths of ASKAP and WSRT with Phased Array Feeds (PAFs) installed, as well as survey areas and redshift ranges for WALLABY, DINGO and WNSHS. The relevant information is reproduced in table \ref{tab:surveys}. WNSHS and WALLABY have quite similar specifications, so we choose to focus on WALLABY here in the knowledge that any findings transfer almost directly to WNSHS. The ambitious depth of LADUMA and CHILES represent somewhat different challenges regarding confusion, from the the medium deep surveys, and we leave the discussion of these to a later paper.

Our theoretical detection limit model (described in the appendix), assuming a signal to noise (S/N) threshold of 5.75, fits very closely to ALFALFA's measured 50\% completeness limit. We assume this form of detection limit for both WALLABY and DINGO, and make use of the properties listed in table \ref{tab:surveys} to estimate confusion in these upcoming surveys.

\begin{table*}
\begin{center}
\begin{tabular}{ | l | c | c | c | c | c | }
\hline
Survey & Area & Resolution & $\sigma_{\mathrm{rms}}$ & Redshift & Time \\ 
Name & ($\mathrm{deg^{2}}$) & & (mJy/15 $\mathrm{km\,s^{-1}}$) & Range & (hr) \\ \hline
HIPASS & 21,350 & 15.5' & 12 & $z <$ 0.04 & 4,300 \\
ALFALFA & $\sim$6,900 & 4' & 2.0 & $z <$ 0.06 & 4,742 \\
WALLABY$^{1}$ & 30,940 & 30'' & 0.81 & $z <$ 0.26 & 9,600 \\
WNSHS$^{1}$ & 10,313 & 13'' & 0.48 & $z <$ 0.26 & 16,900 \\
DINGO$^{1}$ & 150 & 30'' & 0.10 & $z <$ 0.26 & 2,500 \\
\hline
\end{tabular}
\caption{The parameters of current and proposed wide area, blind, HI surveys presented in this table are used throughout this paper to simulate the results of these surveys. $^{1}$Values predicted by \citet{Duffy+2012c} assuming system temperatures of 50 K (although it now seems likely that the final phased array feed systems will fall short of this temperature goal, and thus these numbers will need to be revised). }
\label{tab:surveys}
\end{center}
\end{table*}

\begin{figure*}
\includegraphics[width=\columnwidth]{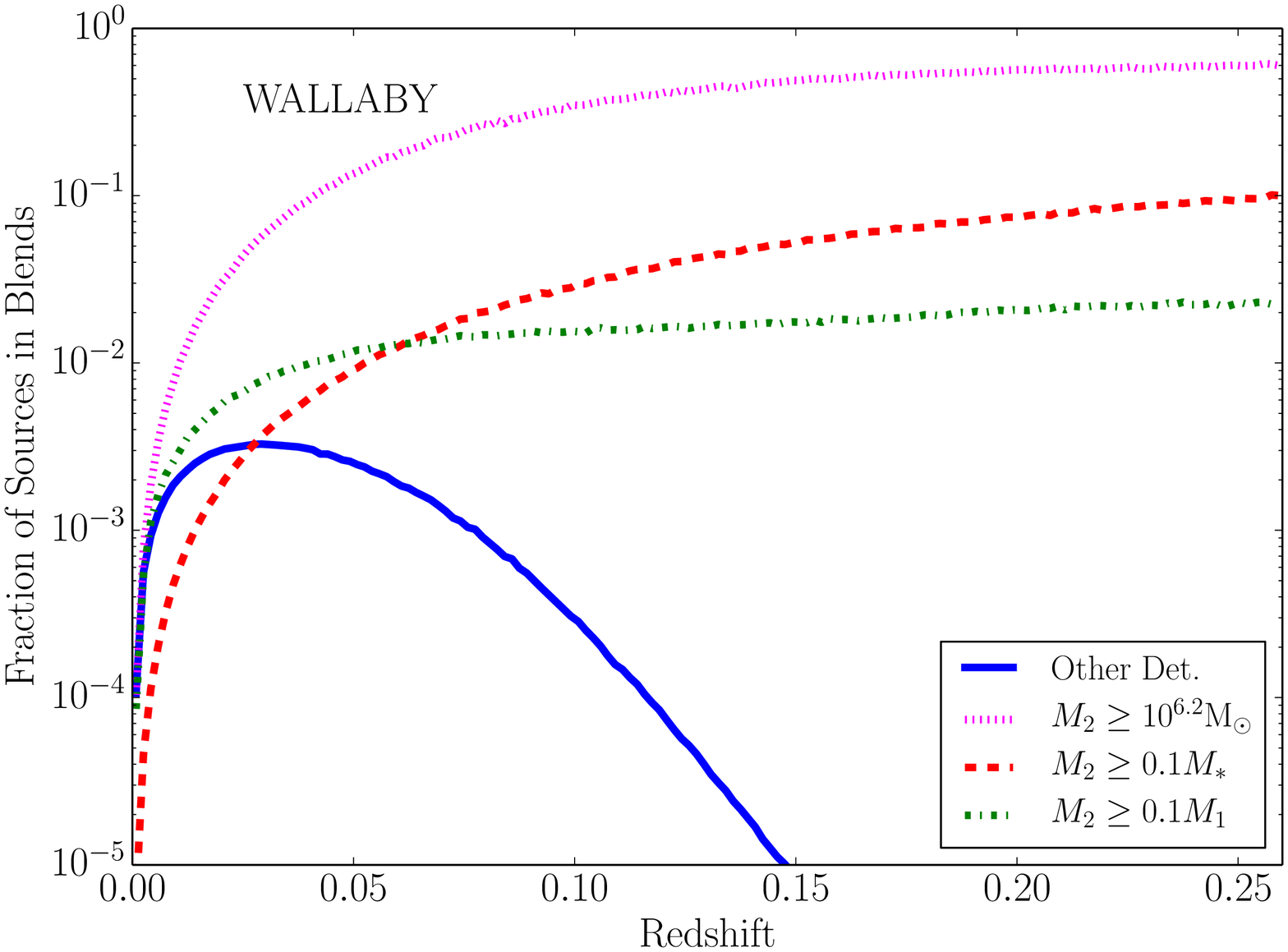}
\includegraphics[width=\columnwidth]{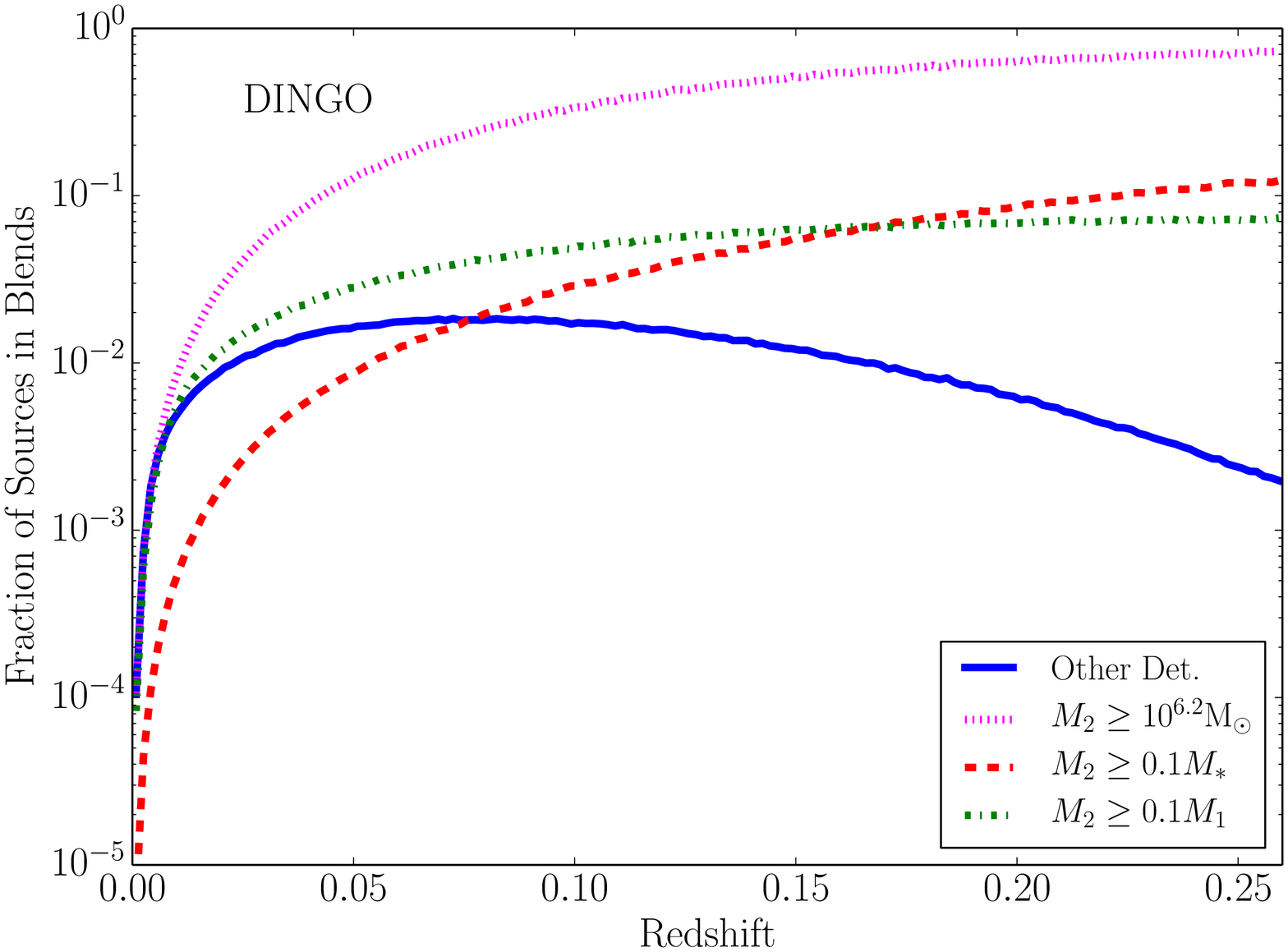}
\caption{Predictions of the rate of confusion in the proposed HI surveys WALLABY (left) and DINGO (right). The blue lines show the rate of confusion with other detected sources (equivalent to figure \ref{fig:AH_conf}), the red dashed lines indicate confusion with galaxies with masses above a tenth of $M_{*}$, the dotted magenta line indicates confusion with any HI galaxy (above an HI mass of $10^{6.2}\;\mathrm{M_{\odot}}$), and the green dash-dot line indicates confusion with any other galaxy above a tenth of the mass of the central galaxy. All of these values lie well below those for ALFALFA or HIPASS.}
\label{fig:WD_conf}
\end{figure*}

Figure \ref{fig:WD_conf} displays four different measures of confusion: confusion with other detections (as plotted above for ALFALFA and HIPASS), confusion with any other HI galaxy (above $10^{6.2} \mathrm{M_{\odot}}$), confusion with HI galaxies that are above a tenth of $M_{*}$ in HI mass, and confusion with other HI galaxies above a tenth of the HI mass of the central galaxy. The first of these represents the amount of confusion that would be apparent in the data, whereas the other three are different measures of the underlying amount of confusion (regardless of detectability). 

At small distances the second definition of confusion is most appropriate, however at large distances where only high mass galaxies are detected this measure is largely irrelevant. Although most galaxies detected at large distance will be blended with at least one other galaxy, that other galaxy will typically be hundreds or thousands of times less massive.

The third measure of confusion is closely related to that used by \DMS, where confusion was defined as the central source being within 30 arcsec (ASKAP synthesised beam), and a fixed velocity range (600 km/s) of another source, that was above 0.1 $M_*$. Using this method they estimated the peak fraction of confused sources in WALLABY and DINGO, would be less than 5\%. This approximation effectively ignores any confusion at lower masses. However as we have already seen, in a survey with little confusion the most noticeable effects occur around the HIMF `knee'. Using our almost equivalent definition of confusion we find the peak fraction to be 10\% and 12\%, for WALLABY and DINGO respectively. \footnotemark The value for DINGO is slightly larger as it can detect galaxies with wider profiles at the same redshift, making blending more probable than in WALLABY.

\footnotetext{The discrepancy between these values and those estimated by \DMS ($\sim$5\%) is due to the combination of a typographical error and potential numerical instability in the solution found in that paper (A. Duffy - private communication), and the different CFs used (although this acts to reduce, rather than increase, our answer). The results reported here have been checked to be stable (see appendix).}

At first glance these numbers may seem to be growing worryingly large, however the equivalent peak value for ALFALFA is $\sim$30\% (note that this is a different measure of confusion to those plotted in figures \ref{fig:AH_conf} \& \ref{fig:conf_m}). Thus, either WALLABY or DINGO would suffer less confusion bias than the currently available large area, blind surveys.

The final measure of confusion is probably the most appropriate for most situations (except when it approaches unity). This measure estimates how frequently a random (detected) galaxy will be blended with something more than a tenth its own mass, and thus potentially introduce a significant error in the measured flux and mass. As it is always significantly below 1, clearly multiple blends are not a concern, even though some of the previous measures may have suggested otherwise. This measure also tends to level out to an almost constant, maximum value beyond a certain redshift. For WALLABY that maximum value is 2\%, and 7\% for DINGO. This indicates that measuring confusion with other sources above 0.1 $M_*$ (red dashed line in figure \ref{fig:WD_conf}), rather than above a tenth the mass of each central source (green dash-dot line in figure \ref{fig:WD_conf}), erroneously implies that WALLABY and DINGO will be equivalently impacted by confusion (10\% and 12\% peak values, respectively). The reason for this is that WALLABY's most distant detections are the most extremely HI-rich galaxies only, where as in DINGO, galaxies near $M_{*}$ are still detectable. This results in their predicted detections being blended at a similar rate with sources above 0.1 $M_*$, but sources above a tenth of the mass of the central source are much more uncommon for WALLABY's most distant detections, than for DINGO's.

As before, this measure of confusion indicates that WALLABY, or any interferometric HI survey of similar depth, will not suffer any global adverse effects due to confusion. DINGO falls in a similar regime to ALFALFA, where confusion is not currently a significant concern, but it would likely become so if the survey were deeper. In addition, one of the aims of DINGO is to measure the evolution of $M_{*}$, and confusion (being a function of redshift also) is likely to be a significant contributor to the error budget of any such measurement.

It should be noted that this analysis is somewhat generous to WALLABY, as it calculates the confusion within one synthesised beam width, whereas $\sim$90\% of its sources will be resolved into at least 2 beams \citep{Duffy+2012c}. However, even if we assume that the beam is actually 1 arcminute across, only 5\% of WALLABY's sources will be confused with galaxies greater than a tenth their own mass, at the outermost redshift where it is likely to detect galaxies (z = 0.15). This value is still multiple times smaller than the equivalent value for ALFALFA or HIPASS.

\begin{figure*}
\includegraphics[width=\columnwidth]{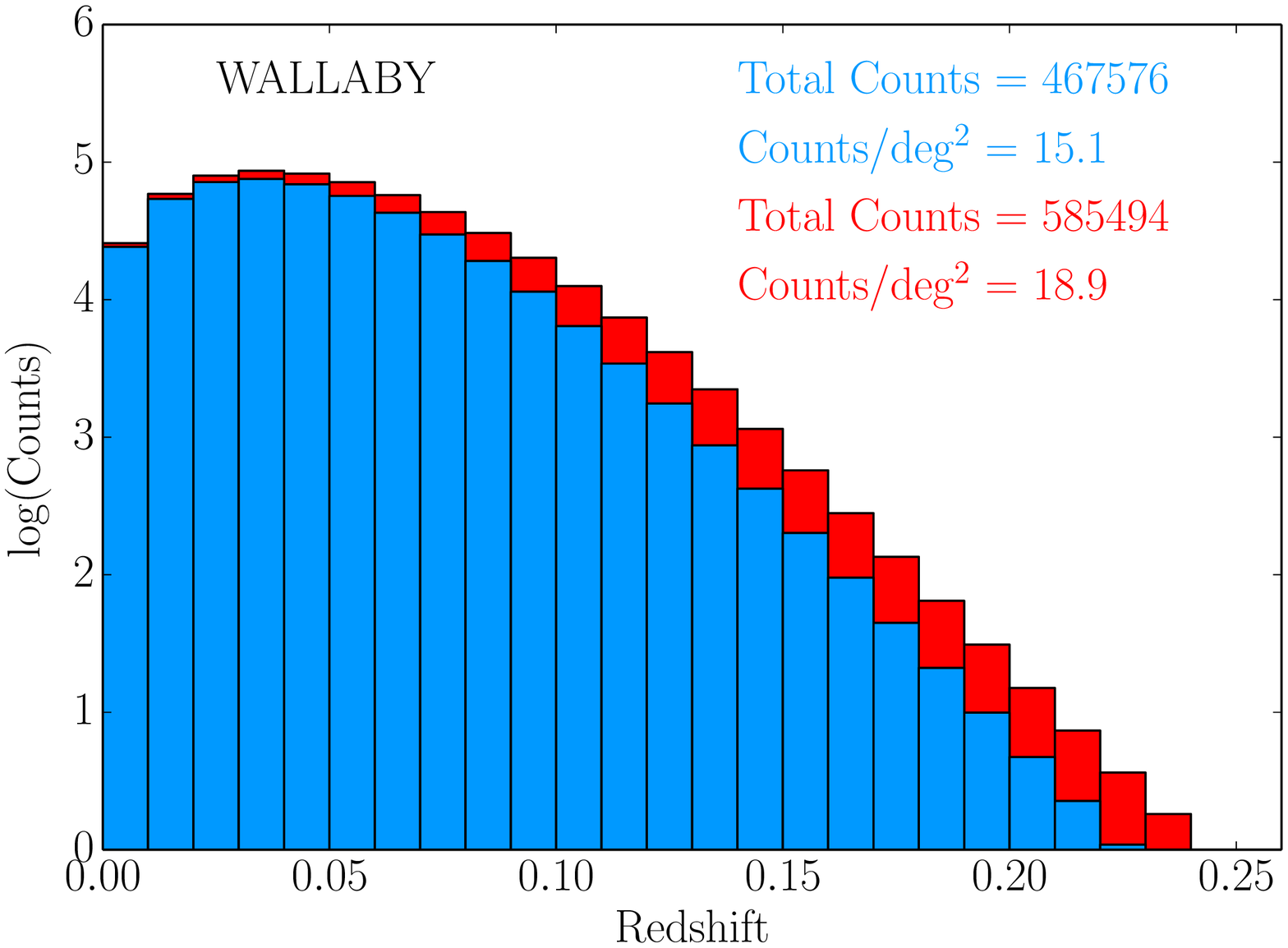}
\includegraphics[width=\columnwidth]{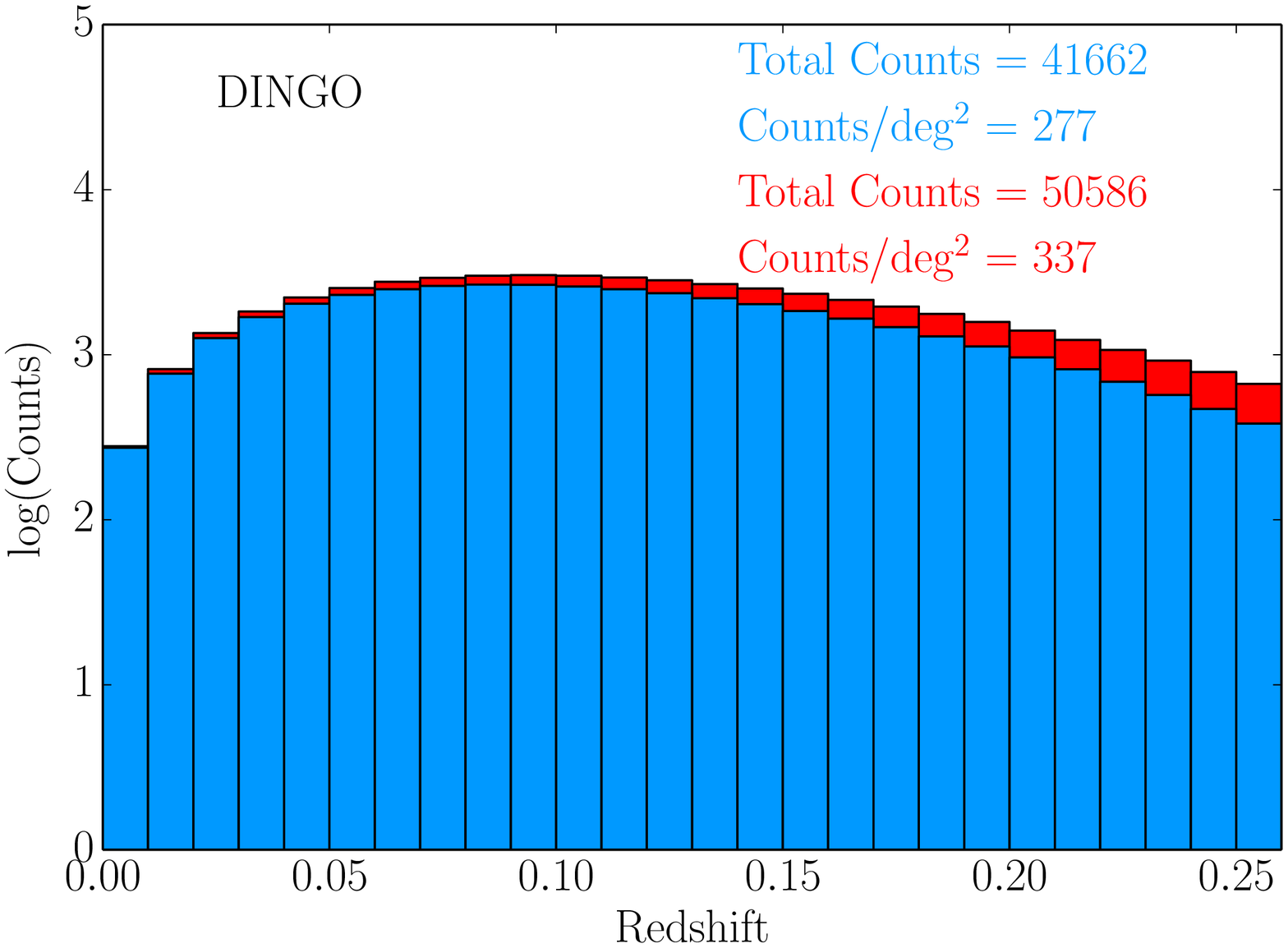}
\caption{Predicted detection number counts of WALLABY (left) and DINGO (right), within redshift bins of width 0.01. The blue bars correspond to a S/N threshold of 5.75 using our detection model, while the red bars assume a straight detection threshold at a S/N of 5 (as in \citet{Duffy+2012c}).}
\label{fig:WD_det}
\end{figure*}
In addition to computing confusion estimates, a byproduct of our model is estimates of the number of galaxies detected as a function of z (assuming no evolution of the HIMF with redshift). Figure \ref{fig:WD_det} shows the predictions for the number of detections WALLABY and DINGO would make. The blue bars show the expected number counts, assuming a source extraction process equivalent to ALFALFA's (modelled as a kinked threshold at 5.75$\sigma$ - see figure \ref{fig:detlim}), and the red bars show the expectation if a straight detection threshold at 5$\sigma$ is used \citep[as in][]{Duffy+2012c}.

This model predicts number counts that are approximately 60\% and 75\% of those estimated in \citet{Duffy+2012c}, for WALLABY and DINGO respectively. Relaxing the detection limit to what was used in that paper only recovers an additional 15\%. The remaining 10-25\% discrepancy must be due to differences between a model based solely on the HIMF (this paper) and one based on populating simulated dark matter halos with HI gas via semi-analytic models. It is not clear which on these is the more reliable approach, however the model presented here accurately reproduces the two currently available wide area, blind HI surveys. However, those surveys are at low redshift and our model does not incorporate any evolution of the HIMF.

In addition, it should be noted that the above discussion entirely neglects the issue of resolving out sources, which \citet{Duffy+2012c} estimate will remove 15\% of WALLABY's sources (though not DINGO's). Finally, as our detection model is based on ALFALFA's pipeline, where every potential source identified by the automated extractor is also examined by hand (a feat that would not be possible for WALLABY, barring a citizen science project), it seems unlikely that WALLABY would be able to match the detection limit assumed here. However, even with all these concerns, WALLABY is still sure to detect more extragalactic HI sources than all current such sources combined.

\subsection{What is the limit of a single dish?}
\label{sec:dish_lim}

Arecibo is the largest single dish telescope in the world, and with the advent of SKA-precursors and new wide area, blind HI surveys, it is appropriate to ask whether single dish telescopes, like Arecibo and the Five hundred metre Aperture Spherical Telescope (currently under construction in China), have a further role to play in this endeavour. One can easily envisage an ALFALFA-like survey that is deeper, however due to the time necessary for such a survey it is likely it would only be carried out if a new 40 beam phased array feed (PAF) were commissioned for the observatory. 

At this point another question becomes relevant: when does the increased confusion, associated with increased depth, prevent accurate measurement of the HIMF with a single dish telescope? To address this question we simulated such surveys with integration times equal to 1, 2, 4, and 8 times that of ALFALFA, but assumed the survey would be truncated at $z=0.05$. For each simulation we calculated a confused and unconfused HIMF, fit Schechter functions to them, and tabulated the deviation in table \ref{tab:deepALFA}.

\begin{table*}
\begin{center}
\begin{tabular}{ | l | c | c | c | c | c | c | }
\hline
Survey & $\Sigma$ & Survey & Survey time & $\Delta \alpha$ & $\Delta m_{*}$ & $z_{\mathrm{max}}$ \\ 
& ($\mathrm{deg^{-2}}$) & time (hr) & w/ PAF40 (hr) & & (dex) & \\ \hline
ALFALFA 1 & $\sim$4 & 4,800 & 840 & -0.03 & 0.06 & 0.05 \\
ALFALFA 2 & $\sim$5 & 9,600 & 1,680 & -0.03 & 0.09 & 0.05 \\
ALFALFA 4 & $\sim$8 & 19,200 & 3,360 & -0.03 & 0.12 & 0.05 \\
ALFALFA 8 & $\sim$11 & 38,400 & 6,720 & -0.01 & 0.15 & 0.05 \\
HIPASS & 0.2 & 4,300 & & -0.04 & 0.07 & 0.04 \\
WALLABY & $\sim$15 & 9,600 & & -0.002 & 0.003 & 0.26 \\
DINGO & $\sim$280 & 2,500 & & -0.007 & 0.02 & 0.26 \\
\hline
\end{tabular}
\caption{The predicted survey timescales and confusion biases for imagined ALFALFA-like surveys with greater integration times (but truncated at z = 0.05) if Arecibo were to with upgraded to a 40 beam PAF. We assume that such a PAF would be cooled to 30 K, and have a sensitivity equivalent to ALFA. Source density on the sky has been denoted as $\Sigma$, and $\Delta \alpha$ and $\Delta m_{*}$ (where $m_{*} = \log M_{*}$) indicate the deviation in the faint end slope and the `knee' mass (in dex) due to confusion. The full HIPASS and the proposed ASKAP surveys are included for comparison. The final column indicates the maximum redshift at which a 21 cm detection could possibly be made given the (assumed) bandwidth. All source density and deviation values (except for HIPASS) assume a detection threshold of 5.75$\sigma$.}
\label{tab:deepALFA}
\end{center}
\end{table*}

Table \ref{tab:deepALFA} shows the estimated completion times for surveys over the ALFALFA sky ($\sim$7000 $\mathrm{deg^{2}}$), if Arecibo were to be upgraded to a 40 beam PAF. The survey names correspond to the factor increase in integration time. The comparison to other surveys is not quite fair as all the ALFALFA-like simulations are truncated at $z=0.05$. For clarity, a depth equivalent to WALLABY occurs between 4 and 8 times the integration time of ALFALFA. The reason for this truncation is twofold: firstly, Puerto Rico has serious RFI concerns beyond a redshift of $\sim$0.05, making accurate determination of the completeness difficult, and secondly because confusion will potentially dominate the uncertainty in $M_{*}$ for a survey deeper than ALFALFA with Arecibo's resolution, thus any such survey must focus on the faint end slope, and the relevant galaxies will not be detected beyond this redshift.

It should be noted that the survey times given in table \ref{tab:deepALFA} correspond to the factor gained due to having 40 beams rather than 7, only. The exact completion time of any such deeper survey would depend on the beam pattern and how the drifts are tiled on the sky. We also assume that a 40 beam PAF at Arecibo would be cooled to 30 K (as is ALFA), whereas the PAFs on ASKAP are assumed to be at 50 K. Cooling PAFs on an interferometer presents a more complex engineering challenge, compared to cooling a similar device on a single dish telescope, as each antenna must be have its own cooling system. As the noise level scales linearly with the system temperature any increase in the assumed temperature will result in the relevant survey losing sensitivity by the same factor, unless its timescale were to be increased by that factor squared.

As can be seen in table \ref{tab:deepALFA} the deviation of the faint end slope ($\alpha$) due to confusion, is not a simple function of survey depth, and in fact is smaller in the simulations of 2 and 4 times the integration time of ALFALFA, than for the original simulation. The reason for this is because we are dealing with a fixed volume. In a fixed volume, as the survey becomes deeper, the galaxies above $M_{*}$ are quickly all detected, thus the mass where the effect of confusion transitions from suppressing a bin to enhancing it, decreases. As the transition point shifts to before the `knee', the deviation of $M_{*}$ stagnates, and confusion begins to lift the more massive end of the faint end slope, flattening, rather than steepening it. Neither of these effects would occur in a survey with unlimited bandwidth.

Although the effects described are expected to occur to some degree, the results of these simulations should be approached which caution. The deviations calculated are intended to be upper limits, but in the fixed volume case they may be sensitive to the simplistic assumption that confusion merely combines the flux (mass) of two objects. This is because the position of the transition point is entirely governed by the relative impact of confusion on adjacent bins. To better understand this, a more realistic model of how the flux of one source blends in to another, and how this influences both the measured flux and velocity width, would be required. Despite this, the general result still stands, that the faint end slope measured by a deeper survey in a fixed volume, is not necessarily more impacted by confusion.

Finally, when considering the extreme of the faint end slope a key advantage of single dish telescopes over interferometers is that they have poor resolution. Almost no extragalactic source will be resolved out by any single dish telescope, regardless of its mass or proximity. This simplifies the statistical corrections required to accurately measure the faint end slope. However, it is at present unclear what impact this effect will have on the ability of surveys like WALLABY and WNSHS to probe very low mass galaxies.

If Arecibo were to focus on a certain region of sky, rather than repeating all the ALFALFA sky, one such volume of interest might be the Pieces-Perseus supercluster (PPS) ridge, spanning a 4$^{\circ}$ strip in declination (from 28$^{\circ}$ to 32$^{\circ}$), between about 22 and 3 hours right ascension. ALFALFA currently has $\sim$900 detections within 9,000 $\mathrm{km\,s^{-1}}$ in this strip, and simulations indicate that a 4 times longer survey would increase this to $\sim$1,500.

In this direction there is a deep foreground void, where ALFALFA only detects tens of galaxies, out to 3,000 $\mathrm{km\,s^{-1}}$. However, the PPS overdensity between 4,000 and 8,000 $\mathrm{km\,s^{-1}}$ is so strong, that the overall surface density of detections in this strip is one and a half times that of the rest of ALFALFA. A deeper map of this volume would thus allow the HIMF to be investigated both in void and supercluster environments, open the door for peculiar velocity studies around these structures (as few redshift have been measured in this region), and create a sample of low mass void galaxies, all with one dataset. Such a survey would require an additional 525 hours with ALFA, or a total of 160 hours with a 40 beam PAF. On the practical side, Arecibo's limited steer-ability and the need for night-time observing would restrict the window for observations to the period between Aug 15th and Dec 1st, and thus such a survey would likely take several years to be executed.

In summary, interferometric surveys aim to trace HI out to greater redshifts, probe any redshift evolution of the high mass end of the HIMF, and will be capable of entering a parameter space that confusion may obscure from single dish telescopes. However, a convincing detection of environmental dependence of the faint end slope has yet to be made, although it is expected from $\Lambda$CDM \citep{Peebles2001,Tinker+Conroy2009}. Thus, if future single dish HI surveys are to remain competitive in this field, they should play to their strengths and focus on studying the environmental dependence of the HIMF (particularly the faint end slope), and nearby, extremely low mass galaxies.

\section{Conclusions}

In general we found that confusion acted to alter the HIMF in the same ways: steepening the faint end slope ($\alpha$) and increasing the `knee' mass ($M_{*}$). The influence of confusion on the shape of the HIMF is non-linear, and can be counter-intuitive. The reason for this is that the shape of a function depends on the relative shifts occurring in adjacent bins, as well as the absolute change, which in turn depend on both the survey resolution and its depth. Meaning that the shape of an HIMF from a more confused survey is not necessarily more impacted by confusion.

We have developed a comprehensive model to describe the rate at which HI sources will be spectroscopically confused in a given survey, as a function of redshift. This model shows good agreement with the observable confusion present in the ALFALFA survey and HIPASS. Our simulations indicate that neither of those surveys have serious biases stemming from confusion, and that, of the differences in their HIMFs, only the faint end slope might be attributed to confusion bias. The upper limits of the alterations to the Schechter function parameters that describe their HIMFs, are placed at 3$\sigma$ (based on published random errors), and in reality could be significantly smaller.

Encouragingly, $\alpha$, was the parameter most resilient against the influence of confusion. Studies searching for environmental dependence of the HIMF by using the ALFALFA and HIPASS datasets should therefore focus on this parameter. Detection of a 3$\sigma$ deviation from the slopes of the published $\alpha$.40 or HIPASS HIMFs would be robust against the effects of confusion, however a similar deviation in $M_{*}$ may not be.

Simulations of proposed medium depth upcoming SKA precursor experiments (WALLABY and DINGO) indicated approximately a factor of 2 more confusion than had previously been predicted, however they would still be less confused than either HIPASS or ALFALFA. For WALLABY the maximum potential bias from confusion was found to be smaller than the random counting errors, and for DINGO it was of the same order as the random errors. Surveys that go deeper than DINGO, but with equivalent resolution, will once again be in the regime of ALFALFA and HIPASS, where a deeper survey with the same telescope will not necessarily return a more accurate HIMF.

Our model also predicts that the ASKAP surveys will detect around 60-75\% of the number of sources that had previously been estimated, however this would still be over an order of magnitude greater than ALFALFA and HIPASS combined. A small fraction of this discrepancy can be explained by the different detection limits assumed, however the bulk of it is likely due to differences between a model based on the mass-width function, and one based on semi-analytic models and halo catalogues.

As in the coming years interferometer based surveys will have far better confusion statistics than single dish surveys, and due to modern phased array feeds, will have vastly improved survey speeds, it begs the question ``where can single dishes still be competitive in surveying extragalactic HI?" Other than projects carrying out HI intensity mapping (a whole other field in itself), the answer likely lies in deeper (but fixed volume) surveys that focus on environmental dependence and the lowest mass galaxies, two fields where much is still to be done. The shallow redshift would prevent excess confusion, allowing studies of the faint end slope to remain robust against confusion, while their lower resolution would prevent systematic biases due to the angular extent of nearby, low mass galaxies; which together would permit single dish telescopes to probe an area of cosmology and galaxy evolution that would be more difficult with any other type of instrument.

\section*{Acknowledgements}

The authors acknowledge the work of the entire ALFALFA collaboration in observing, flagging, and extracting the catalogue of galaxies that this paper makes use of. The ALFALFA team at Cornell is supported by NSF grants AST-0607007 and AST-1107390 to RG and MPH and by grants from the Brinson Foundation. EP is supported by a NOVA postdoctoral fellowship at the Kapteyn Institute. MGJ would like to thank Alan Duffy for a useful discussion concerning the correlation function, as well as his constructive comments as reviewer. We also thank P. Erdo{\v g}du and C. Springob for providing the 2MRS overdensity map, that we used to correct for LSS.

\bibliography{confusion_arxiv}

\appendix

\section{Detection Limit}
\label{sec:detlim}

In order to develop a general expression for the detection threshold of a survey given its predicted rms noise per channel, channel width, and redshift range; we follow \citet{Giovanelli+2005}, which made a prediction for ALFALFA's detection limit, and make changes where appropriate.

The peak flux from an HI source ($\mathrm{Jy}$) can be approximated as
\begin{equation}
\centering
S_{\mathrm{peak}} = \frac{M_{\mathrm{HI}}}{2.356 \times 10^{5} d^{2} W (1+z)} \frac{\mathrm{Mpc^{2} \, km\,s^{-1}}}{\mathrm{M_{\odot}}},
\end{equation}
where $M_{\mathrm{HI}}$ is the HI mass of the galaxy in $M_{\odot}$, $W$ is its velocity width in $\mathrm{km\,s^{-1}}$ (corrected for cosmological expansion), and $d$ is the comoving distance \citep[using WMAP9 cosmology from][]{Hinshaw+2013} to it in $\mathrm{Mpc}$. The factor of $(1+z)^{-1}$ results from competing effects due to the cosmological expansion \citep{Peacock1999,Abdalla+Rawlings2005}.

For a given telescope and frontend one can measure (or model) the system temperature and gain, in order to predict the rms noise per channel. Assuming this number ($S_{\mathrm{rms}}$) is available, the only aspects left to consider are the fraction of the source contained within the beam or synthesised beam ($f_{b}$), and the effect of smoothing to maximise signal to noise (S/N). This leaves us with the following expression for S/N
\begin{equation}
\label{eqn:S/N}
\centering
S/N = \frac{M_{\mathrm{HI}} f_{b} \sqrt{f_{\mathrm{smo}}}}{235.6 \, d^{2} W S_{\mathrm{rms}}} \frac{\mathrm{mJy \, Mpc^{2} \, km\,s^{-1}}}{\mathrm{M_{\odot}}},
\end{equation}
where $f_{\mathrm{smo}}$ is the number of channels that the signal can be smoothed over. Here an additional factor of $(1+z)$ enters, which cancels out the previous factor, due to the fact that a uniformly tiled (or drift scan) survey effectively integrates a given point in the sky for longer at higher redshift, because the beam area grows in proportion to $(1+z)^{2}$, resulting in a factor of $(1+z)$ increase in expected sensitivity. As noted in \citet{Duffy+2012c}, ASKAPs PAFs are designed to maintain approximately constant overlap between synthesised beams, regardless of redshift, which will negate this second effect. Therefore, equation \ref{eqn:S/N} will have a factor of $(1+z)^{-1}$ when considering ASKAP's HI surveys. 

As long as smoothing occurs over regions containing signal, it will give a $\sqrt{f_{\mathrm{smo}}}$ increase to the S/N; as the signal increases linearly with the number of channels smoothed over, but the noise only increase like the square root. However, in practice very broad HI profiles have much less flux at their centre frequency than in the two horns, thus at some point smoothing will give diminishing returns. \citet{Haynes+2011} found that for ALFALFA the transition width ($W_c$) occurs at $\log{W_{c}/\mathrm{km\,s^{-1}}} = 2.5$, and we adopt this value throughout. Thus the maximum number of channels a source can be smoothed over, is just the ratio of the larger of $W$ or $W_{c}$, to the channel width.
\begin{eqnarray}
f_{\mathrm{smo}} = \frac{1}{\Delta v_{\mathrm{ch}}} \left\{
\begin{array}{ll}
W & \mathrm{if} \; W \leq W_{c} \\
W_{c} & \mathrm{if} \; W > W_{c}
\end{array}
\right.
\end{eqnarray}
Where $\Delta v_{\mathrm{ch}}$ is the channel velocity width (at $z=0$). No redshift dependence is included for $\Delta v_{\mathrm{ch}}$ as $W$ is the intrinsic velocity width, that is, it is already corrected for cosmological redshift.

In order to set the threshold value of signal to noise for an HI detection, we compare this model to the 50\% completeness limit found by \citet{Haynes+2011} for the $\alpha.40$ sample, which has an $S_{\mathrm{rms}}$ of 3.4 mJy per 24.4 kHz channel. A signal to noise threshold of 5.75 gives a very close approximation to the measured completeness limit. In practice the completes of any survey will depend on the data reduction and extraction process. As ALFALFA implements both an automated extraction algorithm \citep{Saintonge+2007}, and visually inspects every potential source, it is unlikely that a purely automated process will recover an equivalent threshold, and in this sense it can be considered a lower limit.

We adopt a S/N threshold of 5.75 to simulate ALFALFA's extraction process, and apply this to all other simulation, with the exception of HIPASS, where we used the published completeness surface \citep{Zwaan+2004}. Also for simplicity, we assume $f_{b} = 1$ for all sources within all simulated surveys. This is essentially always true for single dish surveys, but interferometric surveys are likely to resolve a significant fraction of HI galaxies, which will somewhat degrade their detection capabilities.

\section{2D Correlation Function}
\label{sec:2DCF}

The correlation function gives the excess probability (compared to random) that at a given velocity and angular separation from a source, there is another source. To find the probability that a given source will be confused, we need to know what the probability that at least one other source is within a certain projected separation perpendicular to the line-of-sight, $\kappa_{\mathrm{sep}}$ (dependent on the telescope beam, and distance), and velocity separation, $\beta_{\mathrm{sep}}$ (dependent on the velocity widths of the two galaxies). This scenario is best described by an inhomogeneous Poisson process; a Poisson process where the occurrence rate varies with position. Using this framework gives the probability of another source being within $\kappa_{\mathrm{sep}}$ and $\beta_{\mathrm{sep}}$ as
\begin{equation}
\centering
p(\kappa_{\mathrm{nearest}} < \kappa_{\mathrm{sep}} \cap \beta_{\mathrm{nearest}} < \beta_{\mathrm{sep}}) = 1 - e^{-\left< N(\kappa_{\mathrm{sep}},\beta_{\mathrm{sep}}) \right>},
\end{equation}
where subscript `nearest' denotes the values of the central source's nearest neighbour, and $\left< N \right>$ was defined in section \ref{sec:model} as:
\begin{eqnarray}\nonumber
\label{eqn:N_av}
\left< N \right> &=& 2 \int_{W_{\mathrm{min}}}^{W_{\mathrm{max}}} \int_{M_{\mathrm{lim}(d,W_{2})}}^{M_{\mathrm{max}}} \phi(M_{2}) p(W_{2}|M_{2}) \nonumber\\ && \int_{0}^{\frac{W_{1}+W_{2}}{2 H_{0}}} \int_{0}^{\Theta_{\mathrm{beam}} d} 2\pi \kappa \left( 1+\xi(\kappa,\beta) \right) \nonumber \\ && \mathrm{d}\kappa \, \mathrm{d}\beta \, \mathrm{d}M_{2} \, \mathrm{d}W_{2},
\end{eqnarray}
where here $\kappa_{\mathrm{sep}}$ corresponds to $\Theta_{\mathrm{beam}} d$, and $\beta_{\mathrm{sep}}$ is $(W_{1}+W_{2})/2 H_{0}$.

In order to evaluate $\left< N \right>$ we must first fit an expression to the 2D CF \citet{Papastergis+2013}. We take the simplest form that is not axisymmetric, a function that is elliptical in the $\kappa\beta$-plane:
\begin{equation}
\centering
\xi(\kappa,\beta) = \left( \frac{1}{r_{0}}\sqrt{\frac{\kappa^{2}}{a^{2}} + \frac{\beta^{2}}{b^{2}}} \right)^{\gamma},
\end{equation}
where $ab = 1$ and the best fit gives $r_{0} = 9.05$ Mpc, $a = 0.641$, and $\gamma = -1.13$. This fit and the data are shown in figure \ref{fig:2DCF}. This fit demonstrates that there is a slight `finger of god' effect present in the data, as the velocity axis is stretched relative to the angular axis. On scales larger than 10 Mpc, the apparent contraction of structure along the line-of-sight becomes the more obvious effect, however we do not see this in our fit because we only fit the CF for separations smaller than 10 Mpc, as larger separations are not relevant to the study of confusion.

Now to calculate $N(\kappa,\beta)$ we must evaluate the spatial integrals in equation \ref{eqn:N_av}, which gives
\begin{eqnarray}\nonumber
2 \int_{0}^{\beta_{\mathrm{sep}}} \int_{0}^{\kappa_{\mathrm{sep}}} & 2\pi \kappa \left( 1+\xi(\kappa,\beta) \right) \mathrm{d}\kappa \mathrm{d}\beta = \\ 
& 2 \pi a \left[ \frac{\beta_{\mathrm{sep}} \kappa^{2}_{\mathrm{sep}}}{b a^{2}} + I \right],
\end{eqnarray}
where
\begin{eqnarray}\nonumber
&&I = \frac{2\frac{\beta_{\mathrm{sep}}}{b} \left( \frac{\kappa_{\mathrm{sep}}}{a} \right)^{\gamma+2}(\gamma+3)}{(\gamma+2)(\gamma+3) r_{0}^{\gamma}} \\ &&\left[ _{2}F_{1}\left( \frac{1}{2},-\frac{\gamma}{2}-1;\frac{3}{2};-\frac{a^{2} \beta^{2}_{\mathrm{sep}}}{b^{2} \kappa^{2}_{\mathrm{sep}}} \right) -2\left( \frac{\beta_{\mathrm{sep}}}{b} \right)^{\gamma+3} \right]
\end{eqnarray}
and $\mathrm{_{2}F_{1}}$ is the Gaussian hypergeometric function. A similar solution to this integral was derived in \DMS, however that solution was found to be unstable over the relevant parameter space. The solution above was compared against numerical integration for a range of physical parameters and gave consistent results in all cases.

\section{Conditional Velocity Width Function}

Once the HI mass of a given galaxy, and its position relative to its neighbours, has been determined via the HIMF and the CF, its velocity width must also be determined before it is possible to assess whether it is involved in a spectroscopic blend with a neighbour. To calculate the mass conditional velocity width function (MCWF) we follow a similar approach to \citet{Martin+2010}, where a Gumbel distribution is fit to the velocity width distribution within narrow mass bins, however here we weight each data point by $1/V_{\mathrm{eff}}$ \citep[see][]{Zwaan+2005,Papastergis+2011}. The trend in the parameters of the Gumbel fits is then modeled to produce a simple analytic expression for the probability of a galaxy of mass $10^{m}\;\mathrm{M_{\odot}}$ having a velocity width $10^{w}\;\mathrm{km\,s^{-1}}$.
\begin{equation}
\centering
p(w|m) = \frac{1}{\beta(m)} \frac{e^{-\left( z(m) + e^{-z(m)} \right)}}{e^{-e^{-z_{\mathrm{min}}}} - e^{-e^{-z_{\mathrm{max}}}}},
\end{equation}
where $z = \frac{\mu(m) - w}{\beta(m)}$, $z_{\mathrm{min}}$ and $z_{\mathrm{max}}$ correspond to the minimum and maximum allowed values of $w$, $\mu$ is the distribution center, and $\beta$ is its width, which are given by
\begin{equation}
\centering
\mu = 0.322m - 0.728
\end{equation}
and
\begin{eqnarray}
\beta = \left\{
\begin{array}{ll}
-0.0158m + 0.316 & \mathrm{if} \: m \leq 9.83 \\
-0.0578m + 0.729 & \mathrm{if} \: m > 9.83
\end{array}
\right.
\end{eqnarray}
Additionally the above distribution is only valid for $\log 15 < w < 3$, and is set to zero beyond these to prevent the production of unphysical velocity widths.

\end{document}